\documentclass[journal]{IEEEtran}
\usepackage[pdftex]{graphicx}
\usepackage[cmex10]{amsmath}
\usepackage{amssymb}
\usepackage{multicol}
\usepackage{colortbl}%
  
\usepackage{longtable}
\usepackage{caption}
\usepackage{subcaption}
\usepackage{cite}
\usepackage{psfrag}
\usepackage{epstopdf}
\usepackage{algorithm}
\usepackage{algorithmic}
\usepackage{widetext}
\usepackage{color}
\usepackage{soul}
\usepackage{upgreek}
\usepackage{multirow}
\usepackage[justification=centerlast]{caption}
\usepackage{tikz,pgfplots}
\usepackage{dblfloatfix}
\usepackage{textcomp}
\usepackage{url}
\usepackage{adjustbox}
\usepackage{algorithm}
\usepackage{algorithmic}
\usepackage{amsmath}
\DeclareUnicodeCharacter{2005}{\hspace{0.01em}}
\usetikzlibrary{calc}
\usepackage{mathtools}
\usetikzlibrary{arrows}
\pgfplotsset{compat=newest}
\usepgfplotslibrary{fillbetween}
\usetikzlibrary{patterns}

\usepackage[hidelinks]{hyperref}
\hypersetup{breaklinks=true}
\usepackage{cite}

\usepackage{multicol}
\usepackage{colortbl}%
\usepackage{longtable}

\renewcommand{\arraystretch}{1.2}
\newcommand*\GG[1]{\textcolor{gray}{~[~\textbf{GG: $\rightarrow$}~#1~]}}
\newcommand*\Pedro[1]{\textcolor{blue}{~[~\textbf{Pedro: $\rightarrow$}~#1~]}}

\newcommand*\J[1]{\textcolor{green}{~[~\textbf{Jar $\rightarrow$}~#1~]}}
\newcommand{\AMD}{AMD\textsuperscript{\textregistered}}

\definecolor{myBlue}{RGB}{72,125,215}
\definecolor{myOrange}{RGB}{118,54,45}
\definecolor{InfinBlue}{RGB}{72,72,51}

\ifCLASSINFOpdf
\else
\fi
\begin{document}
% \title{ Theory and Hardware Implementation of Time-Domain Clustered Equalizer for Chromatic Dispersion Compensation in Coherent Optical Links}

\title{Geometric Clustering for Hardware-Efficient Implementation of Chromatic Dispersion Compensation }
     \pgfplotsset{
        % use this `compat` level or higher to make use of the "advanced"
        % label positioning (this brings the second ylabel to the right)
        compat=1.3, 
        % (created a style for the common options)
        my axis style/.style={
            every axis plot post/.style={/pgf/number format/fixed},
            ybar=5pt,
            bar width=8pt,
            x=1.7cm,
            axis on top,
            enlarge x limits=0.1,
            symbolic x coords={MLP, biLSTM, ESN, CNN+MLP, CNN+biLSTM, DBP},
            %restrict y to domain*=0:1200, % Cut values off at 14
            visualization depends on=rawy\as\rawy, % Save the unclipped values
%            after end axis/.code={ % Draw line indicating break
%                \draw [ultra thick, white, decoration={snake, amplitude=1pt}, decorate] (rel axis cs:0,1.05) -- (rel axis cs:1,1.05);
%            },
            nodes near coords={%
                \pgfmathprintnumber[precision=2]{\rawy}% Print unclipped values
            },
            every node near coord/.append style={rotate=90, anchor=west},
            tick label style={font=\footnotesize},
            xtick distance=1,
        },
    }

\author{Geraldo Gomes, Pedro Freire, Jaroslaw E. Prilepsky, Sergei K. Turitsyn
\thanks{SKT acknowledges the EPSRC project TRANSNET (EP/R035342/1).}
\thanks{Geraldo Gomes, Pedro Freire, Jaroslaw E. Prilepsky  and Sergei K. Turitsyn are with Aston Institute of Photonic Technologies, Aston University, United Kingdom, p.freiredecarvalhosourza@aston.ac.uk.}
% \thanks{ Bernhard Spinnler are with Infinera R\&D, Sankt-Martin-Str. 76, 81541, Munich, Germany. anapoli@infinera.com.}

\thanks{Manuscript received xxx 19, zzz; revised January 11, yyy.}}


\markboth{Journal of Lightwave technology , ~Vol.~y, No.~x, November~2023}%
{Shell \MakeLowercase{\textit{et al.}}:  xxxxxxxxxxxx}



\maketitle
\begin{abstract}
% Power efficiency is one of the key challenges in modern optical fiber communication systems, leading, in particular, to the continuous efforts to decrease the computational complexity of digital signal processing and, specifically, of chromatic dispersion compensation (CDC) algorithms. A fair comparison of the existing approaches for the CDC still lacks hardware implementation for some of the proposed solutions to comprehensively validate the benefits. This paper provides a theoretical analysis of the tap overlapping effect in CDC filters for coherent receivers and, based on that, introduces a novel Time-Domain Clustered Equalizer (TDCE) technique. An innovative way to perform parallelization of TDCE was developed and used for the hardware implementation of this equalizer for fiber lengths up to 640 km. A fair comparison with the state-of-the-art frequency domain equalizer (FDE) under the same conditions is provided. We demonstrate that the implementation strategy and factors like parallelization and memory implementation are no less crucial then the computational complexity in determining the overall complexity at the hardware level and energy efficiency. We demonstrate that the hardware implementation of the proposed TDCE results in up to 70.7\% energy savings and 71.4\% multiplier usage savings compared to FDE despite having a higher computational complexity.

Power efficiency remains a significant challenge in modern optical fiber communication systems, driving efforts to reduce the computational complexity of digital signal processing, particularly in chromatic dispersion compensation (CDC) algorithms. While various strategies for complexity reduction have been proposed, many lack the necessary hardware implementation to validate their benefits. This paper provides a theoretical analysis of the tap overlapping effect in CDC filters for coherent receivers, introduces a novel Time-Domain Clustered Equalizer (TDCE) technique based on this concept, and presents a Field-Programmable Gate Array (FPGA) implementation for validation. We developed an innovative parallelization method for TDCE, implementing it in hardware for fiber lengths up to 640 km. A fair comparison with the state-of-the-art frequency domain equalizer (FDE) under identical conditions is also conducted. Our findings highlight that implementation strategies, including parallelization and memory management, are as crucial as computational complexity in determining hardware complexity and energy efficiency. The proposed TDCE hardware implementation achieves up to 70.7\% energy savings and 71.4\% multiplier usage savings compared to FDE, despite its higher computational complexity.

\end{abstract}

% Note that keywords are not RMpSally used for peerreview papers.

% \fi
%
% For peerreview papers, this IEEEtran command inserts a page break an
\IEEEpeerreviewmaketitle
% \newpage
\section{Introduction}

% \IEEEPARstart{E}{nergy} consumption has been a keen interest subject in optical transmission systems for a relatively long time\cite{agrell2016roadmap,radovic2024power,freire2023low}. In pluggable optical coherent transceivers, digital signal processing (DSP) typically accounts for about 50\% of the total power usage\cite{minkenberg2021co}. One of the main power-consuming components within DSP is the CDC\cite{kuschnerov2014energy}. Because of this, recent proposals, such as chirp filtering \cite{felipe2020chirp} and uniform quantization of filter taps\cite{martins2016distributive,wu2022low}, have been proposed, showing considerable computational complexity savings but lacking the hardware implementation validation to confirm the benefits in terms of energy efficiency and chip area. Other designs have been implemented in hardware, such as the finite field approach\cite{ji2023hardware}. However, the aforementioned work assessed the resulting efficiency only for short-reach fiber systems (240km). The comparison to the state-of-the-art was performed by using scaling factors and assuming that the impact of different parameters, like throughput and technology process size, grows linearly.

\IEEEPARstart{E}{nergy} consumption has long been a critical focus in optical transmission systems\cite{agrell2016roadmap,radovic2024power,freire2023low, freire2023tutorial}. In pluggable optical coherent transceivers, digital signal processing (DSP) typically accounts for about 50\% of total power usage\cite{minkenberg2021co,nagarajan2021low}, with the CDC being one of the primary power-consuming components\cite{kuschnerov2014energy}. Recent proposals, such as chirp filtering \cite{felipe2020chirp} and uniform quantization of filter taps\cite{martins2016distributive,wu2022low}, have shown significant computational complexity reductions but lack validation through hardware implementation to confirm benefits in energy efficiency and chip area. Some designs, like the finite field approach\cite{ji2023hardware}, have been implemented in hardware. However, these efforts were focused on short-reach fiber systems (240 km) and relied on scaling factors, assuming a linear growth in the impact of different parameters like throughput and technology process size, for comparison to the state-of-the-art.

This paper provides a theoretical analysis of the tap redundancy in CDC filters for coherent receivers. It presents different approaches to decrease the complexity of time domain equalization based on this phenomenon with the associated trade-offs. We introduce a novel approach called the Time-Domain Clustered Equalizer (TDCE), which leverages the tap redundancy phenomenon to reduce the complexity of the time-domain equalizer. Moreover, the proposed technique is suitable for optimization using machine learning, allowing us to gain further reduction of the complexity. Our equalizer is implemented in a Field-Programmable Gate Array (FPGA) for different fiber lengths up to 640 km. These implementations are compared, under the same conditions, against the frequency domain equalizer (FDE)\cite{xu2010coherent} implementation using Fast Fourier Transform (FFT) algorithm\cite{cooley1965algorithm}, as the state-of-the-art because the latter has been widely studied\cite{bae2021finite,bae2023fft} and implemented in Application-Specific Integrated Circuit (ASIC) for CDC\cite{sun2020800g,fougstedt2020asic}. We evaluate chip area usage, hardware resources, and energy efficiency using nJ/recovered-bit for both solutions.

Importantly, we show that the implementation strategy plays a vital role
(comparable with the computational complexity) in the analysis of the efficiency of the physical resources utilization.  In particular, we emphasize that two factors are critically important: (i) the innovative parallelization of TDCE and (ii) the memory implementation. It is demonstrated that a lower multiplier count is possible for a higher complexity algorithm due to the hardware implementation strategy utilized. Our analysis illustrates that the computational complexity metric alone is insufficient to determine the hardware characteristics of the equalizer. It is necessary to examine other crucial points, such as memory implementation and the degree of parallelization the algorithm allows. Furthermore, in agreement with Ref.~\cite{kovalev2017implementation}, we show that theoretical multiplication complexity is not a good predictor for the energy efficiency of CDC filters.

In summary, our work provides a hardware implementation of the TDCE design, demonstrating its effectiveness in reducing power consumption and multiplier usage across various fiber lengths compared to FDE. Despite higher theoretical computational complexity, our results highlight the significant impact of hardware implementation strategies on energy consumption and resource efficiency.

\section{Theoretical Analysis}

\subsection{Filter taps overlapping}

There have been several works dealing with the CDC that employed clustering of time-domain filter taps in IM/DD pre-compensation scenarios, arguing the tap redundancy~\cite{huang2023low, xie2023low,xie2024deployment}, though without a theoretic explanation. Our paper deals with a different problem -- the CDC for coherent detection systems, which requires a complex filter that has different types of redundancy. This section will explain the redundancy effect.

To begin the explanation, we reiterate the well-known observation that each filter tap represents a point on a circle in the complex plane~\cite{taylor2008compact}. It is because each filter tap is given by \cite{savory2008digital}:
\begin{equation}
    g\left[m\right] = \sqrt{\frac{jcT^2}{D\lambda^2 z}} \exp\left(-j \frac{\pi c T^2}{D\lambda^2 z}m^2\right),
    \label{taps-equation}
\end{equation}
in which, \(z\) is the fiber length, \(m\) is the filter tap index,  \(j\) is the imaginary unit, \(D\) is the dispersion coefficient, \(c\) is the speed of light, \(\lambda\) is the wavelength of the carrier, and \(T\) is the sampling period.
The boundaries for \(m\) are calculated as follows \cite{savory2008digital}:
\begin{equation}
    -\left\lfloor \dfrac{N}{2} \right\rfloor \leq m \leq 
    \left\lfloor \dfrac{N}{2} \right\rfloor,
    N = 2 \left\lfloor \dfrac{|D|\lambda^2 z}{2cT^2} \right\rfloor+1,
    \label{taps-bound}
\end{equation}
in which, \(N\) is the maximum number of filter taps, taking into consideration the Nyquist frequency to avoid aliasing, and $\left\lfloor x \right\rfloor$ means the nearest integer less than $x$. As $m$ can assume negative values according to Eq.~(\ref{taps-bound}), we prefer to use in this paper the index $k$ given by:
\begin{equation}
k=m+\left\lfloor \dfrac{N}{2} \right\rfloor, k=0,...,N-1,
\end{equation}
\iffalse
\J{explain the meaning of brackets for N/2 -- an integer part?}
\GG{Done}

\J{Why do you discuss Eq 4 here? Discuss it after Eq 4}
\GG{Done}

\J{Why $N$ is not explained?}
\GG{Done}

\fi

\begin{figure}[htbp]
        \centering
        \includegraphics[height=3.92cm]{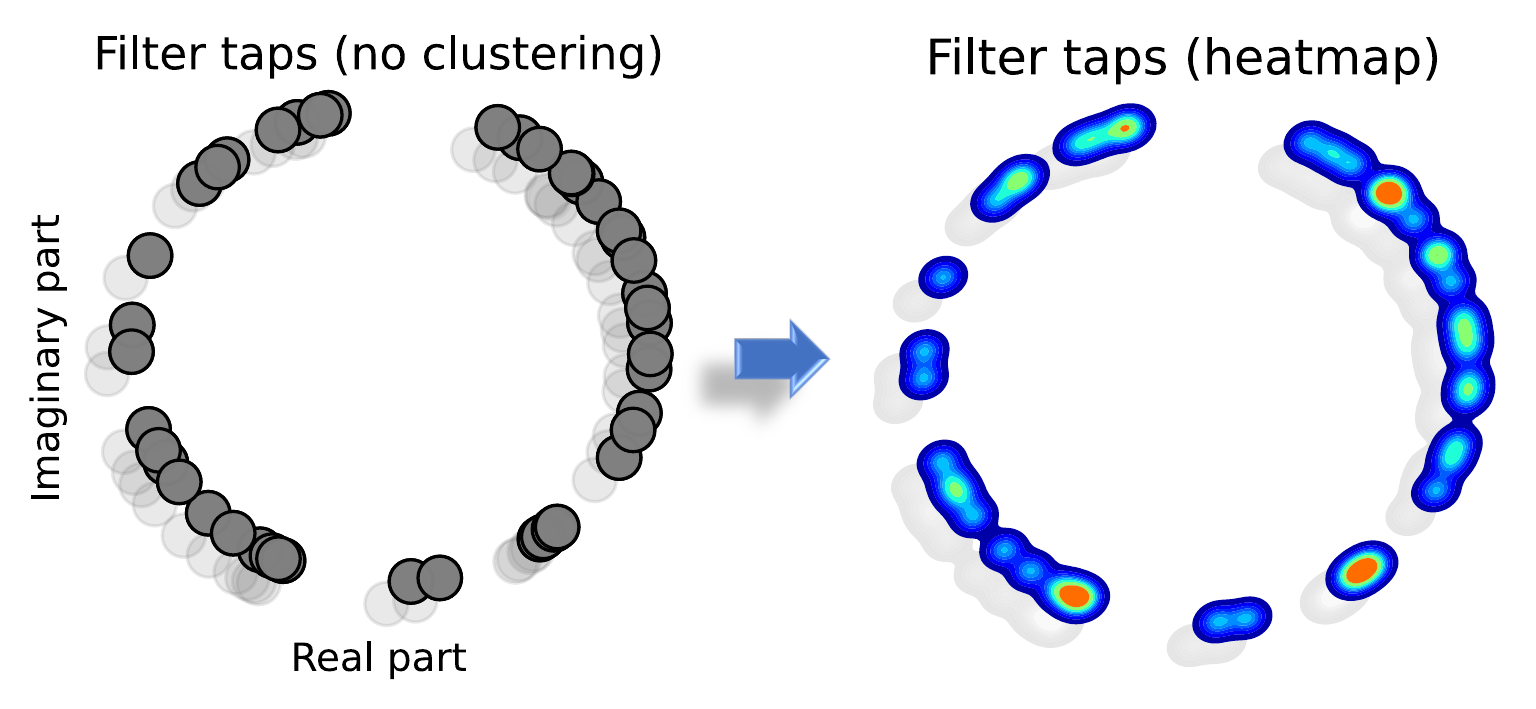}
        \caption{Tap redundancy illustration. On the left, each gray circle represents a filter tap in the complex plane, showing clearly the tap accumulation. On the right, a heat map shows spots in areas of high concentration of filter taps. Filter taps calculated for 80km, 32Gbaud, 2 samples/symbol, dispersion coeff. $D$ = 16.8 ps/(nm$\cdot$km) and $\lambda$ = 1550nm}
        \label{fig:cluster-motivation}
    \end{figure}

In Fig.~\ref{fig:cluster-motivation}, the filter taps calculated by Eq.~(\ref{taps-equation}) are plotted in the complex plane (left) and it is possible to observe the overlapping of filter taps, while in (right) a heat-map with brighter spots in areas of high accumulation of taps illustrate the formation of clusters.
\iffalse
\J{I do not see ``possible clusters'' in the right fig. Rephrase for clarity}
\GG{Done}
\fi

% \iffalse \J{explain the meaning of brackets for N/2 -- an integer part?} \GG{moved N to next equation} \fi which represents a rotating vector with constant amplitude and varying phase as tap index $m$ varies. Therefore, a multiplication by a filter tap can be seen as simply a phase shift of the input sample with a constant scaling factor at the end. \J{Why do you discuss Eq 4 here? Discuss it after Eq 4} In Eq.~(\ref{taps-equation}), \(z\) is the fiber length, \(m\) is the filter tap index,  $N$ is the maximum number of filter taps taking into consideration the Nyquist frequency to avoid aliasing, \(j\) is the imaginary unit, \(D\) is the dispersion coefficient, \(c\) is the speed of light, \(\lambda\) is the wavelength of the carrier, \(T\) is the sampling period, and $\left\lfloor x \right\rfloor$ means the nearest integer less than x. \iffalse \J{Why $N$ is not explained?} \GG{Explained now.} \fi

This cluster formation can be explained by the fact that the filter tap definition given by Eq.~(\ref{taps-equation}) represents a rotating vector with constant amplitude and varying phase as tap index $k$ varies. Therefore, a multiplication by a filter tap can be seen as simply a phase shift of the input sample with a constant scaling factor at the end. As the absolute phase increases for different values of $k$, many phases will be repeated (because the even symmetry due to the \(m^2\) term) or have values very close to each other on the complex plane circle because the values greater than \(2\pi\) just represent multiple rounds on the complex circle.

In this paper, we show how the analysis of the clusters of phase shifts makes it possible to reduce the complexity of the time domain equalizer through the utilization of factorization of the finite impulse response (FIR) filtering process, executing first the sums of the samples associated with filter taps that are grouped and then multiplying the sum result by the filter tap that represents the group, which we call the clustered filter tap.

% \begin{figure}[htbp]
% \centering
% \setlength{\arrayrulewidth}{0pt} % Set the width of the table borders to zero
% \begin{minipage}{0.45\linewidth}
% \centering
% \includegraphics[width=\linewidth]{rose_plot_80km.pdf}
% \end{minipage}%
% \begin{minipage}{0.45\linewidth}
% \centering
% \includegraphics[width=\linewidth]{rose_plot_8000km.pdf}
% \end{minipage}
% \caption{Circular histograms for different distances using 30bins (each bin is 12°) considering 32Gbaud, dispersion coeff. $D$ = 16.8 ps/(nm$\cdot$km) and $\lambda$ = 1550nm }
% \label{fig:rose-plots} % Label for the figure
% \end{figure}

\begin{figure}[htbp]
\centering
\setlength{\arrayrulewidth}{0pt} % Set the width of the table borders to zero
\centering
\begin{tabular}{|c|c|c|}
\hline
\textbf{80km, 27 taps} & \textbf{8000km, 2647 taps} \\
\hline
\includegraphics[width=0.45\linewidth]{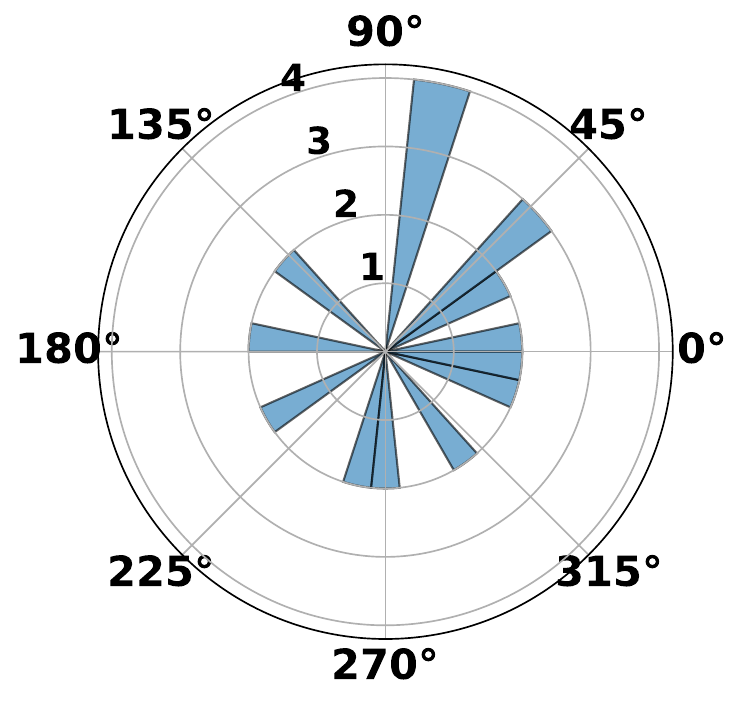} & \includegraphics[width=0.45\linewidth]{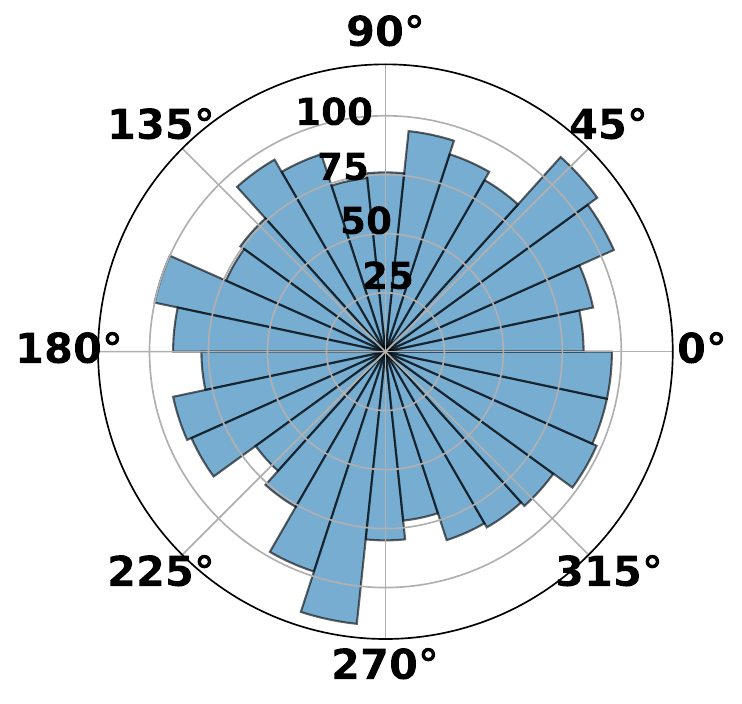} \\
\hline
\end{tabular}
\caption{Circular histograms showing how many filter taps are present in each bin (30 bins in total, 12°/bin). On the left, a clear clustered distribution of the filter taps is shown for the 80km equalizer filter, whilst on the right, a uniform-like distribution is shown for a higher dispersion scenario (8000km). Both equalizer filters are calculated for 32Gbaud, 2 samples/symbol, $D$ = 16.8 ps/(nm$\cdot$km) and $\lambda$ = 1550nm }
\label{fig:rose-plots}
\end{figure}

To investigate further this overlapping, in Fig.~\ref{fig:rose-plots} we plot circular histograms to statistically analyze the angle distribution for equalizing filters in two different scenarios. The filter size utilized \(M\) was 60\% of the maximum number of taps \(N\) in Eq.~(\ref{taps-bound}), which gives a considerable reduction in the filter size with minimal penalties~\cite{savory2008digital}.

Fig.~\ref{fig:rose-plots} reveals two distinct patterns: for short-range distances, we see a cluster formation (also illustrated in Fig~\ref{fig:cluster-motivation}), and for long-haul distances, the angle distribution is more spread but with some more occupied clusters (e.g., near 45° and 270°).

To assess how concentrated in clusters or evenly spread the filter taps are for different fiber lengths, let us introduce a simple metric \(\rho\),
\begin{equation}
    \rho = \frac{N_{\text{M}}-N_{\text{L}}}{\mu},
    \label{rho}
\end{equation}
which is the difference between the number of taps in the most (\(N_{\text{M}}\)) and lowest (\(N_{\text{L}}\)) populated bins divided by the mean quantity of taps per bin \(\mu\), which is the total number of taps $M$ divided by the total number of bins. In an ideal uniform distribution, all the bins should be populated equally, therefore \(N_{\text{M}}=N_{\text{L}}\) and \(\rho=0\). However, if the difference $N_M-N_L$ is bigger in comparison to $\mu$, it means that at least one bin or more contains a higher concentration of taps than the others.
\iffalse
\J{Mean of what? quantity for points in clusters? Unclear}
\GG{Done}

\J{this explanation "to express" is redundant if you introduce your quantity clearer} 
\GG{Done}

 \J{Why Fig 3 is here? It must be placed after it is mentioned in the text}
 \GG{Done}

\fi

\begin{figure}[htbp]
        \centering
        \includegraphics[height=5cm]{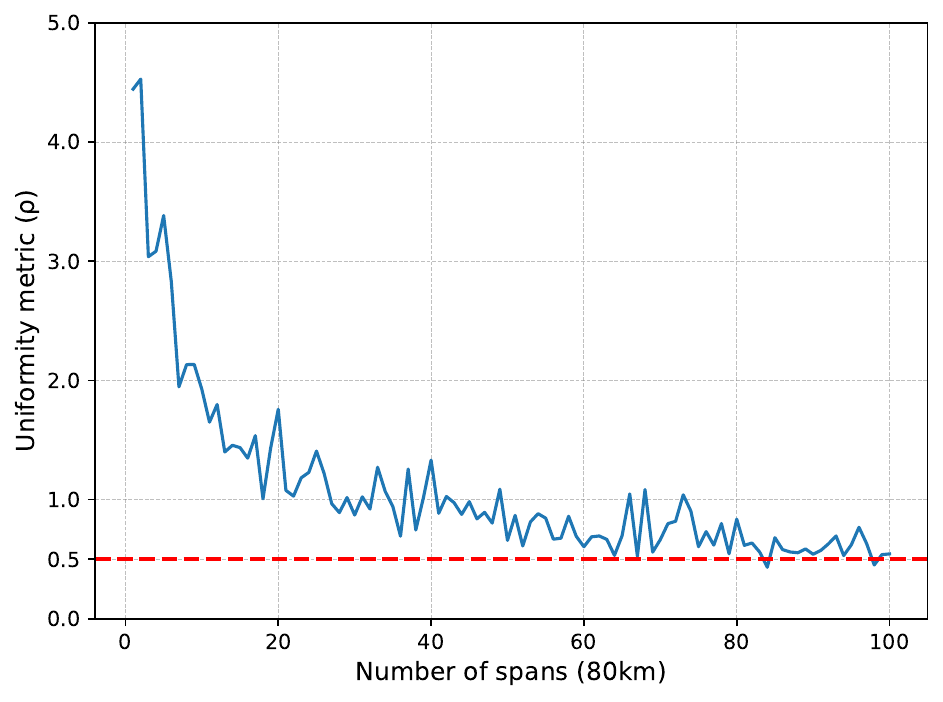}
        \caption{Uniformity analysis, showing that as the dispersion increases (in this case fixed 32Gbaud and increasing distance) the difference between the number of taps in the most populated and least populated bins becomes less significant (\(\rho \leq 1\)) in comparison to the average value (\(\mu\)). This means that there are still clusters, but for high dispersion scenarios, they are not highly localized, depicting a more uniform distribution.}
        \label{fig:uniformity-plot}
    \end{figure}

Looking at Fig.~\ref{fig:uniformity-plot}, one can conclude that for shorter distances, the distribution has one or more points of concentration (clusters), while as the distance increases, the angle distribution becomes more uniform and \(\rho\) values decrease. 
\iffalse
\J{Not a good sentence, rework}
\GG{Done}
\fi

Taking \(\rho=0.5\) as a reference, we can conclude that even for higher distances, which have a more uniform distribution, there is at least one cluster in the distribution. However, for these distances the concentration points are not so prominent. This result interpretation is in line with plots of Fig.~\ref{fig:rose-plots}.

\subsection{Complexity reduction by cluster analysis}
%As in FDE, the complexity is governed by the FFT/IFFT performed~\cite{spinnler2010equalizer,leibrich2010frequency}, the tap overlapping described in the last section seems useless for complexity reduction. However, 

\iffalse

\J{I would delete everything preceding this comment -- not really needed, some emotions} 
\GG{Done}

\fi

Looking at the time domain convolution equation,
\begin{equation}
\label{convolution}
y[n] = \sum_{k=0}^{M-1} x[k] \cdot g[n - k],
\end{equation}
where $M$ is the filter size, we observe that the filter taps \(g[n-k]\) are directly multiplied by the input samples \(x[k]\) to produce the output samples \(y[n]\). Thus, if some filter taps are grouped in clusters, each cluster can be represented by a single tap. 
This approximation can help decrease the complexity by using the distributive property of multiplication: we can first sum all the input samples associated with the grouped filter taps and then multiply this summed result by the tap representing the cluster. A similar use of this distributive property was employed to utilize the even symmetry of the transfer function~\cite{spinnler2010equalizer,martins2016distributive} and uniform quantization~\cite{martins2016distributive,wu2022low}, although none of those works harnessed the geometrical interpretation exploited here.

In this subsection, we explain and calculate the computational complexity, in terms of real multiplications per recovered symbol, of the resulting clustered filter obtained by considering the accumulation of angles in the complex plane, which we call complex-value (CV) clustering.

The FIR filtering by using Eq.~(\ref{convolution}) can simplified by CV clustering and expressed as a simplified convolution:
\begin{equation}
\label{clustered-convolution}
y[n] = \sum_{k=0}^{N_{C}-1} x_{S}[k] \cdot g_{C}[k].
\end{equation}
in which $n$ is the output sample index, \(g_{C}\) represents the clustered filter taps to be multiplied by \(x_{S}\), which represents the summation of input samples that for each $n$ are associated with the same cluster of filter taps, and \(N_{C}\) is the total number of complex value clusters.

\iffalse
\J{A monstrous complicated sentence - split}
\GG{Done}

\J{A formula without a sentence}
\GG{Done}

\J{Defined by a set? Who defined them via a set?}
\GG{Done}
\fi

The index of the input samples that must be summed before the multiplication by a specific clustered filter tap can be defined by a set \(Q_{j}\) in which \(j\) is the cluster's index varying from \(0\) to \((N_{C}-1)\). 
Therefore, the summation of associated input samples can be represented by:
\begin{equation}
\label{clustered-summation}
x_{S}[j] = \sum_{k \in Q_{j}} x[k].
\end{equation}
These sets \(Q_{j}\) play the same role as the routing block described in Ref.~\cite{martins2016distributive}.

Defining the complexity ($C$) as the number of real multiplications per recovered sample and noting that each complex multiplication is made up of 4 real multiplications, this strategy leads to the expression for the complex clustering complexity: $C_{CV} = 4\cdot N_{C}$.

\begin{figure*}[h!]
\centering
% \hspace{-1mm}
\setlength{\arrayrulewidth}{0pt} % Set the width of the table borders to zero
\begin{tabular}{c}
% First row with three columns
\begin{tabular}{ccc}
\begin{tikzpicture}[scale=0.56]
            \begin{axis} [
                xlabel={Number of Taps ($M$)},
                xlabel style={yshift=-5pt,font=\fontsize{13}{14}\selectfont},
                ylabel={\textbf{BER(x1000)}},
                grid=both,  
                ylabel near ticks,
                ylabel style={yshift=0pt,xshift=0pt,font=\fontsize{13}{14}\selectfont},
                yticklabel style={
                    /pgf/number format/.style={
                        fixed,
                        fixed zerofill,
                        precision=0,
                        scientific auto,
                        tick label style={/pgf/number format/1000 sep=,}
                    },font=\large
                },
                xmin=89, xmax=109,
            	xtick={89, 91, 93, 95, 97, 99, 101, 103, 105, 107, 109},
            	ymin=0, ymax=5,
                ytick={1,2,3,3.8},
                legend pos=north east,
                legend style={at={(0.95,0.7)}, 
                legend cell align=left,
                fill=white, 
                fill opacity=0.6, 
                draw opacity=1,
                text opacity=1,
                % legend columns=-1
                },
                grid style={dashed},
                xticklabel style ={font=\fontsize{13}{14}\selectfont},
                height=6.5cm, % Set the height of the plot
                width=9cm]
                % ]

            \addplot[color=blue, mark=square, very thick]     coordinates {
% (177, 0.02765655517578125)
% (175, 0.02765655517578125)
% (173, 0.0247955322265625)
% (171, 0.0286102294921875)
% (169, 0.026702880859375)
% (167, 0.02765655517578125)
% (165, 0.02765655517578125)
% (163, 0.026702880859375)
% (161, 0.02574920654296875)
% (159, 0.02574920654296875)
% (157, 0.02765655517578125)
% (155, 0.02956390380859375)
% (153, 0.02574920654296875)
% (151, 0.0286102294921875)
% (149, 0.02765655517578125)
% (147, 0.026702880859375)
% (145, 0.0286102294921875)
% (143, 0.02765655517578125)
% (141, 0.0286102294921875)
% (139, 0.02765655517578125)
% (137, 0.0247955322265625)
% (135, 0.0247955322265625)
% (133, 0.0286102294921875)
% (131, 0.02956390380859375)
% (129, 0.02765655517578125)
% (127, 0.0286102294921875)
% (125, 0.0286102294921875)
% (123, 0.02574920654296875)
% (121, 0.030517578125)
% (119, 0.03337860107421875)
% (117, 0.034332275390625)
% (115, 0.03719329833984375)
% (113, 0.04482269287109375)
% (111, 0.05340576171875)
(109, 0.05626678466796875)
(107, 0.10013580322265625)
(105, 0.1201629638671875)
(103, 0.19073486328125)
(101, 0.3108978271484375)
(99, 0.5035400390625)
(97, 0.8764266967773438)
(95, 1.5840530395507812)
(93, 2.7618408203125)
(91, 4.6939849853515625)
(89, 7.8220367431640625)
            };
\addlegendentry{Data Points}

    \addplot[dashed, color=green, line width=2pt] coordinates {
        (89, 3.8) (109, 3.8)
    };
    \addlegendentry{Thresh. (FDE)}
    % \addlegendimage{dashed, thick, color=green}
    % \addlegendentry{Thresh. (FDE)}
    
    \addplot[dashed, color=red, line width=2pt] coordinates {
        (89, 1) (109, 1)
    };
    \addlegendentry{Thresh. (TDCE)}
    
                \end{axis}
                \node[text width=3cm] at (1.4,4.9) 
            {\textcolor{red}{\textbf{(a)}}};
\end{tikzpicture}
&
\begin{tikzpicture}[scale=0.56]
            \begin{axis} [
                xlabel={Number of Clusters ($N_{C}$)},
                xlabel style={yshift=-5pt,font=\fontsize{13}{14}\selectfont},
                ylabel={\textbf{BER(x1000)}},
                grid=both,  
                ylabel near ticks,
                ylabel style={yshift=0pt,xshift=0pt,font=\fontsize{13}{14}\selectfont},
                yticklabel style={
                    /pgf/number format/.style={
                        fixed,
                        fixed zerofill,
                        precision=0,
                        scientific auto,
                        tick label style={/pgf/number format/1000 sep=,}
                    },font=\large
                },
                xmin=7, xmax=14,
            	xtick={7,8,9,10,11,12,13,14},
            	ymin=0, ymax=7,
                ytick={1,2,3,3.8,5,6,7},
                legend pos=north east,
                legend style={at={(0.95,0.9)}, 
                legend cell align=left,
                fill=white, 
                fill opacity=0.6, 
                draw opacity=1,
                text opacity=1,
                % legend columns=-1
                },
                grid style={dashed},
                xticklabel style ={font=\fontsize{13}{14}\selectfont},
                height=6.5cm, % Set the height of the plot
                width=9cm]
                % ]

            \addplot[color=blue, mark=square, very thick]     coordinates {
                
                (6, 44.57664489746094)
                (7, 18.136024475097656)
 
                (8,5.8612823486328125)
                (9,4.445075988769531)
                (10,3.71551513671875)
                (11,2.2325515747070312)
 
                (12, 1.2340545654296875)
                (13, 0.4291534423828125)
                (14, 0.1983642578125)

            };
\addlegendentry{Data Points}

    % \addlegendimage{dashed, thick, color=green}
    % \addlegendentry{Thresh. (FDE)}
    
    \addplot[dashed, color=red, line width=2pt] coordinates {
        (7, 3.8) (14, 3.8)
    };
    \addlegendentry{Thresh. (TDCE KNN)}
    
                \end{axis}
                \node[text width=3cm] at (1.4,4.9) 
            {\textcolor{red}{\textbf{(b)}}};
\end{tikzpicture}
&
\begin{tikzpicture}[scale=0.56]
            \begin{axis} [
                xlabel={$\log_2 N_{FFT}$},
                xlabel style={yshift=-5pt,font=\fontsize{14}{14}\selectfont},
                ylabel={\textbf{Real mult./Sample}},
                grid=both,  
                ylabel near ticks,
                ylabel style={yshift=0pt,xshift=0pt,font=\fontsize{13}{14}\selectfont},
                yticklabel style={
                    /pgf/number format/.style={
                        fixed,
                        fixed zerofill,
                        precision=0,
                        scientific auto,
                        tick label style={/pgf/number format/1000 sep=,}
                    },font=\large
                },
                xmin=7, xmax=12,
            	xtick={8,9,10,11,12},
            	ymin=20, ymax=100,
                legend pos=north east,
                legend style={at={(0.95,0.8)}, 
                legend cell align=left,
                fill=white, 
                fill opacity=0.6, 
                draw opacity=1,
                text opacity=1,
                % legend columns=-1
                },
                grid style={dashed},
                xticklabel style ={font=\fontsize{13}{14}\selectfont},
                height=6.5cm, % Set the height of the plot
                width=9cm]
                % ]

            \addplot[color=blue, mark=square, very thick]     coordinates {
% (128,157.54)
% (256,59.84)
% (512,49.95)
% (1024,48.87)
% (2048,50.52)
% (4096,53.33)

(7,157.54)
(8,59.84)
(9,49.95)
(10,48.87)
(11,50.52)
(12,53.33)
            };
\addlegendentry{Radix-2}

            \addplot[color=red, mark=square, very thick]     coordinates {
% (256,46.55)
% (1024,37.76)
% (4096,41.02)
(8,46.55)
(10,37.76)
(12,41.02)
            };
\addlegendentry{Radix-4}
    
                \end{axis}
                \node[text width=3cm] at (1,4.9) 
            {\textcolor{red}{\textbf{(c)}}};
\end{tikzpicture}
\end{tabular}
\\
% Second row with four columns
\begin{tabular}{ccc}
\begin{tikzpicture}
    \node[anchor=south west, inner sep=0] (image) at (0,0) {\includegraphics[width=0.25\linewidth]{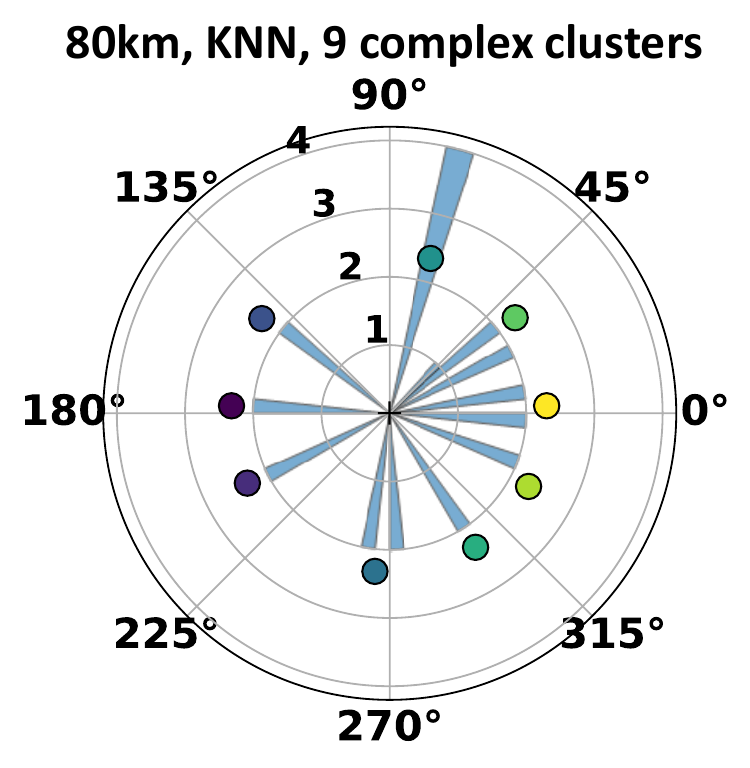}};
    \node at (0,4.2) {\textcolor{red}{\textbf{(d)}}};
\end{tikzpicture}
% &
% \begin{tikzpicture}
%     \node[anchor=south west, inner sep=0] (image) at (0,0) {\includegraphics[width=0.2\linewidth]{cluster-over-histogram-3.pdf}};
%     \node at (0,3.5) {\textcolor{red}{\textbf{(e)}}};
% \end{tikzpicture}
&
\begin{tikzpicture}
    \node[anchor=south west, inner sep=0] (image) at (0,0) {\includegraphics[width=0.25\linewidth]{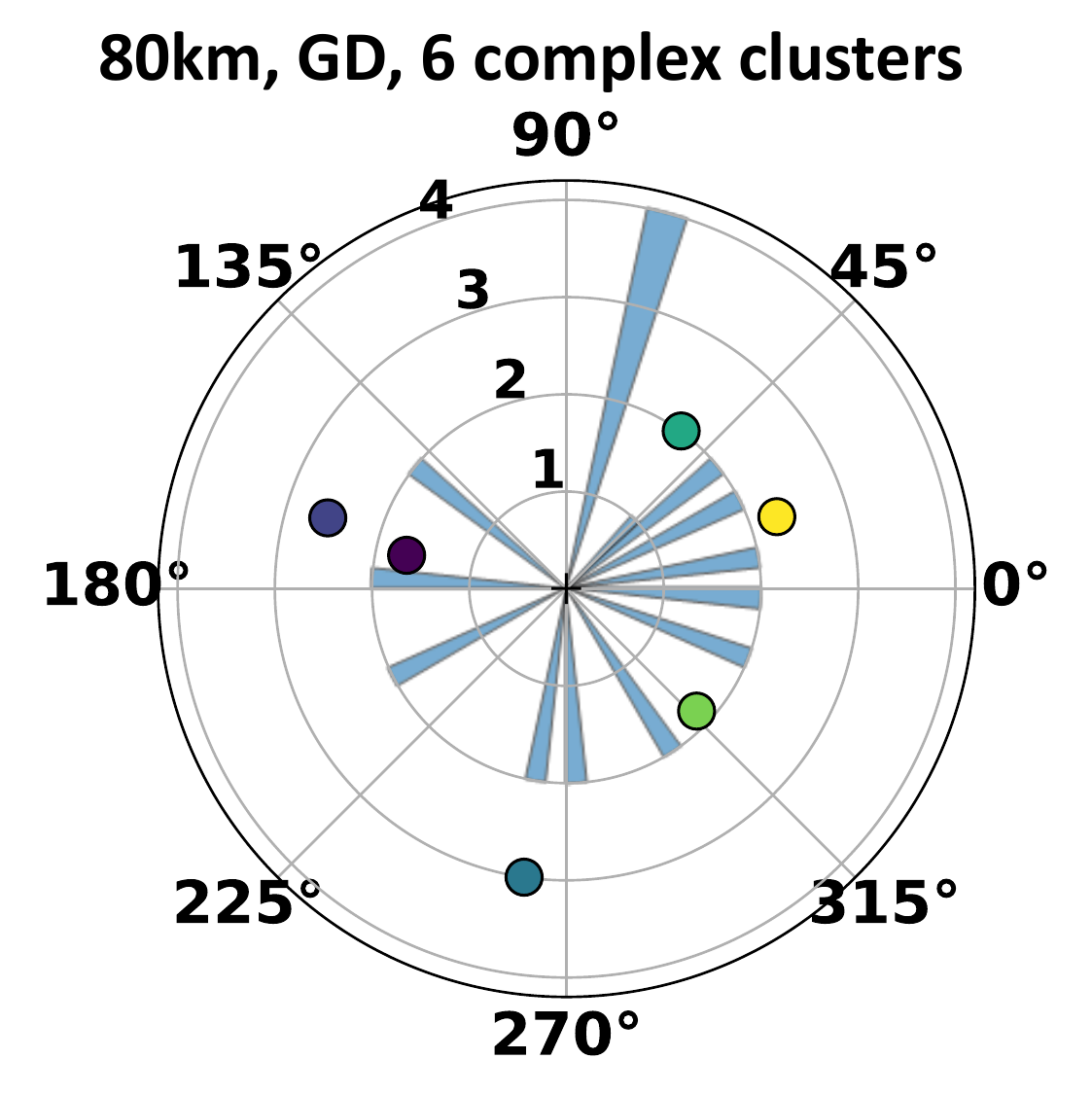}};
    \node at (0,4.2) {\textcolor{red}{\textbf{(e)}}};
\end{tikzpicture}
&
% \vspace{4mm}
\begin{tikzpicture}[scale=0.56]
            \begin{axis} [
                xlabel={Number of Spans (80km)},
                xlabel style={yshift=-5pt,font=\fontsize{13}{14}\selectfont},
                ylabel={\textbf{Real Mult./Sample}},
                grid=both,  
                ylabel near ticks,
                ylabel style={yshift=0pt,xshift=0pt,font=\fontsize{13}{14}\selectfont},
                yticklabel style={
                    /pgf/number format/.style={
                        fixed,
                        fixed zerofill,
                        precision=0,
                        scientific auto,
                        tick label style={/pgf/number format/1000 sep=,}
                    },font=\large
                },
                xmin=1, xmax=8,
            	xtick={0, ..., 8},
            	ymin=20, ymax=55,
                legend style={at={(0.95,1)}, 
                legend cell align=left,
                fill=white, 
                fill opacity=0.6, 
                draw opacity=1,
                text opacity=1,
                legend columns=-1},
                grid style={dashed},
                xticklabel style ={font=\fontsize{13}{14}\selectfont},
                height=8cm, % Set the height of the plot
                width=9cm]
                % ]

            \addplot[color=blue, mark=square, very thick]     coordinates {
            (1, 24)(2, 32)(4, 32)(8, 48) %0.85,1.00,1.47,2.45
            };
            \addlegendentry{TDCE GD};
            
            \addplot[color=red, mark=square, very thick]     coordinates {
            (1, 36)(2, 40)(4, 40)(8, 48) %0.85,1.00,1.47,2.45
            };
            \addlegendentry{TDCE KNN};
            
            \addplot[color=green, mark=*, very thick]   
            coordinates {
            (1, 31.44)(2, 34.46)(4, 37.36)(8, 41.25) %31.44, 34.46, 37.36, 41.25
            };
            \addlegendentry{FDE};
            
                \end{axis}
                \node[text width=3cm] at (1.2,6) 
            {\textcolor{red}{\textbf{(f)}}};

\end{tikzpicture}
\end{tabular}
\end{tabular}
\caption{\footnotesize Methodology used to find the clusters quantities and FFT sizes (top row) and comparison of different clustering approaches (bottom row). a)Truncated filter response for different filter sizes with thresholds utilized for FFT filter and for clustered filter before KNN b) Clustered filter performance for different cluster quantities with pre-FEC threshold c) FDE complexity for different sizes and architectures to find the optimum FFT size d) Cluster location found by KNN in complex plane matching the cluster location visually inferred in the circular histogram (9 complex clusters - circular distribution) e) Cluster location found by KNN applied in the complex plane and optimized by GD algorithm f) Complexity comparison plot between FDE and different clustering approaches for different distances.}
\label{fig:methodology-clustering-methods}
\end{figure*}
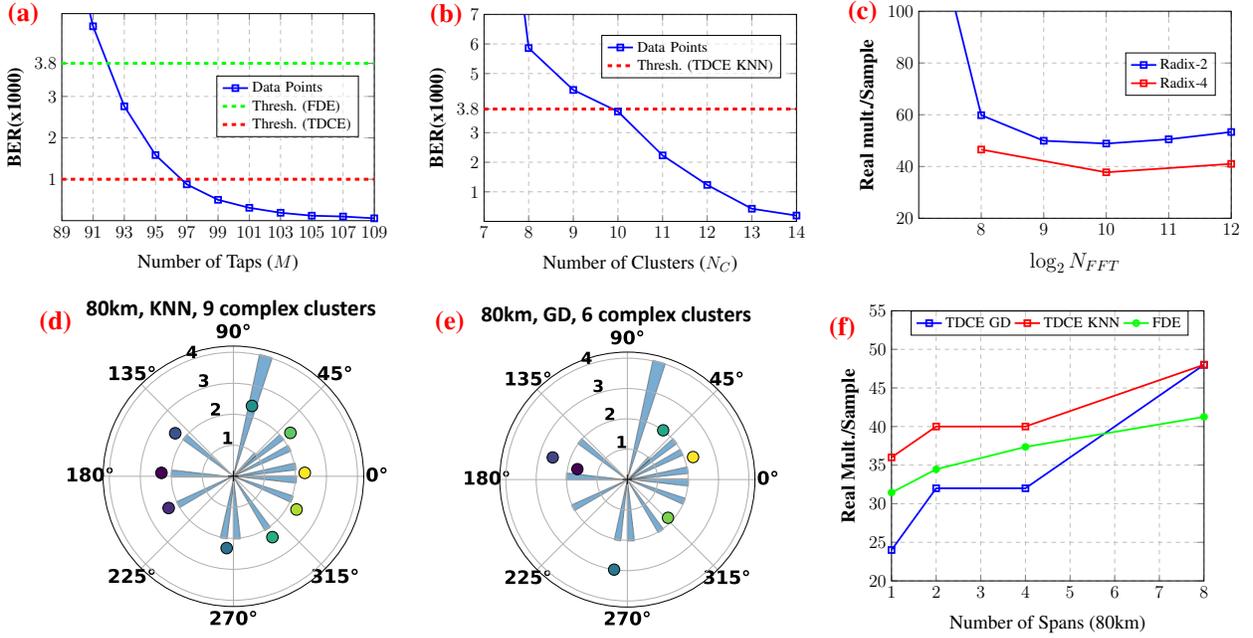

\iffalse
\J{What "pre-summed?"}
\GG{Done}

\J{Amother monstrous and unclear sentence with some unclear formulas inside - rewrite}
\GG{Done}

\J{strategy of what?}
\GG{Done}

\J{At each axis? and the statement is unclear} \Pedro{true the title of the section is complex versus real value clustering you need to continue with the same narrative}
\GG{Done}
\fi
As we aim to demonstrate that, in hardware design, memory usage plays a significant role in both area usage and power consumption, we calculate here the minimum quantity of different memory positions needed to obtain the previously summed values $x_S$ (that we will also refer as pre-summed values). To calculate this, we observe that for each multiplication by $g_C$ in Eq.~(\ref{clustered-convolution}), the real and imaginary values of $x_{S}[k]$ must be calculated previously and saved in memory, which means that $2\cdot N_C$ memory positions are necessary to store $x_S$.

\section{Methodology}\label{sec:methodology}
%

%
\iffalse
\Pedro{ I would rewrite this to :  In this work, we will compare two variations of our approach with the state-of-the-art FDE.
}
\GG{Done}

\Pedro{I would clearly state each approach individually like KNN only and KNN + GD finetuning} 
\GG{Done}

\Pedro{Be careful not to declare acronyms multiple times, as in this case where (GD) appears more than once in the text.}
\GG{Done}
\fi

In this work, we compare two variations of our approach with the state-of-the-art FDE. To find the clusters of filter taps in the complex plane, both TDCE variations use the unsupervised KNN algorithm~\cite{zhang2016introduction}, which has been successful in clustering and reducing equalization complexity in other scenarios~\cite{huang2023low, xie2023low, xie2024deployment}. One variation, TDCE GD, employs supervised learning with the Gradient Descent algorithm (GD) to further optimize the clusters to decrease the approximation errors caused by the clustering. The second variation, TDCE KNN, does not use supervised learning. The methodology for both approaches will be explained individually, considering that TDCE KNN methodology also applies to TDCE GD, with the latter including additional fine-tuning.
% \J{with the "plus"? Do not understand. The whole sentence is dubious - rewrite}

\subsection{TDCE KNN. Methodology}%\label{Sec:low-complex-methods}

\begin{figure*}[h!]
    \centering
    % \hspace{-5mm}
    \includegraphics[width = 0.9\linewidth]{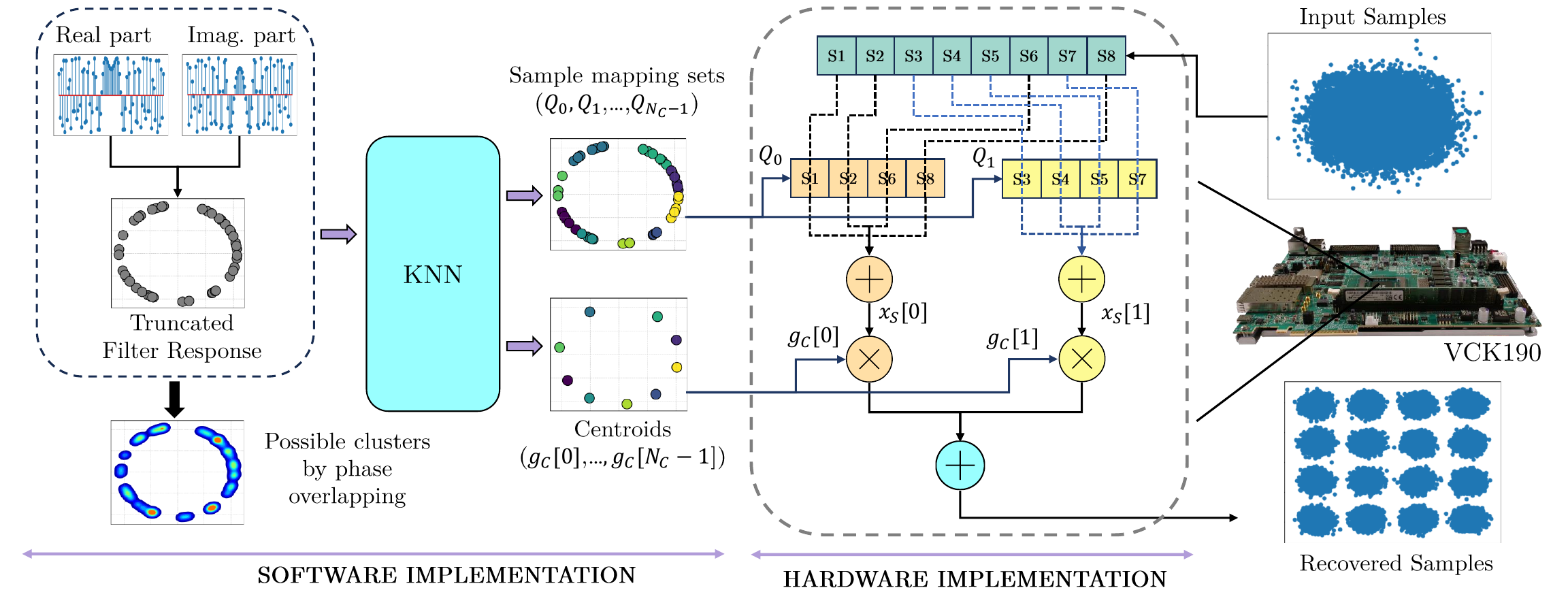}
    \caption{Proposed approach to find the clusters in the transfer function known a priori and how to use this information to achieve a low complexity hardware implementation. On the left side, the real and imaginary parts of the complex transfer function, given by Eq.~(\ref{taps-equation}), are analyzed in the complex plane using the KNN algorithm to generate the clustered filter taps $g_C$ (centroids), and the sample mapping sets $Q$. On the right, sample mappings are used to direct the input samples to the right group of samples that will be summed together to create the pre-summed $x_S$ array before the multiplication by the clustered filter taps (centroids) in the FPGA implementation.}
    \label{fig:cluster-fpga}
\end{figure*}

The clustering process is depicted on the left side of Fig.~\ref{fig:cluster-fpga}, showing the software implementation part. This involves using prior knowledge of variables $T$, $D$, $\lambda$, and $z$ to calculate the truncated filter response $g[k]$ via Eq.~(\ref{taps-equation}). Subsequently, the KNN algorithm identifies the mapping sets $Q_j$ and centroids for the clustered filter taps $g_C[k]$. The right side of Fig.~\ref{fig:cluster-fpga} illustrates the hardware implementation of clustered filtering. This involves routing the input samples to the correct group based on the sample mappings $Q_j$, performing the summation of the samples in the same group to obtain $x_{S}[k]$, and at the end, multiplying the pre-summed values $x_{S}[k]$ with clustered filter taps $g_{C}[k]$, and summing the outputs to get the recovered the symbol.

%\J{didn't understand this}

\iffalse

\Pedro{ Please emphasizes that for fairness reasons, this was also done for the FDE achieving the simplest version for that target BER }
\GG{Done}

\fi
Our objective is to reach a $BER<3.8\times10^{-3}$ in the final TDCE and FDE filters, corresponding to the error-free pre-FEC threshold with 7\% overhead\cite{agrell2018information}. However, overall performance is compromised due to the inherent approximations in the clustering process. Therefore, we conducted an evaluation of the truncated filter response, Eq.~(\ref{taps-equation}), across various filter sizes, ranging from $N$ given by Eq.~(\ref{taps-bound}), to the minimum size that guarantees a BER below $1\times10^{-3}$, as depicted in Fig.~\ref{fig:methodology-clustering-methods}(a). Although this performance is better than our goal, the clustering process will decrease it. Following this, we evaluated the impact of different cluster quantities for each TDCE variation to determine the minimum number of clusters, ensuring a $BER<3.8\times10^{-3}$ as shown in Fig.~\ref{fig:methodology-clustering-methods}(b) for TDCE KNN. A similar approach, explained in detail in the sub-section~\ref{subsec:methodology-fde}, is used for a fair comparison with the minimum complexity FDE.

To assess the filter performances, we utilized data obtained from numerical modeling of a single-channel, 32~GBaud, 16-QAM dual-polarization transmission at the optimal launch power across standard single-mode fiber (SSMF). The propagation dynamic was simulated employing the Manakov equation and the split-step Fourier method \cite{agrawal2000nonlinear}. The SSMF parameters used were: dispersion $D = 16.8$ ps/(nm$\cdot$km), nonlinearity coefficient $\gamma = 1.2$ (W$\cdot$km)$^{-1}$, and attenuation $\alpha = 0.21$ dB/km. Additionally, erbium-doped fiber amplifiers (with a noise figure of 4.5 dB) were included after each fiber span of 80 km. The number of spans ranged from 1 to 8, and all filters operated at 2 samples per symbol.

% \begin{figure}[htbp]
% \centering
% \setlength{\arrayrulewidth}{0pt} % Set the width of the table borders to zero
% \begin{tabular}{cc}
% \begin{tikzpicture}
%     \node[anchor=south west, inner sep=0] (image) at (0,0) {\includegraphics[width=0.45\linewidth]{size-selection.pdf}};
%     \node at (1.4,2.8) {\textbf{(a)}};
% \end{tikzpicture}
% &
% \begin{tikzpicture}
%     \node[anchor=south west, inner sep=0] (image) at (0,0) {\includegraphics[width=0.45\linewidth]{clusters-size-selection.pdf}};
%     \node at (1.4,2.8) {\textbf{(b)}};
% \end{tikzpicture}
% \\
% \begin{tikzpicture}
%     \node[anchor=south west, inner sep=0] (image) at (0,0) {\includegraphics[width=0.45\linewidth]{fft-size-selection.pdf}};
%     \node at (1.4,3) {\textbf{(c)}};
% \end{tikzpicture}
% &
% \begin{tikzpicture}
%     \node[anchor=south west, inner sep=0] (image) at (0,0) {\includegraphics[width=0.45\linewidth]{complexity_comparison.pdf}};
%     \node at (1.4,3) {\textbf{(d)}};
% \end{tikzpicture}
% \end{tabular}
% \caption{\footnotesize a)Truncated filter response for different filter sizes with thresholds utilized b) Clustered filter performance for different cluster quantities c)FFT size optimization plots d) Complexity comparison plot between FDE and TDCE for different distances}
% \label{fig:clustering-process}
% \end{figure}

\subsection{TDCE GD (fine-tuning). Methodology}%\label{Sec:low-complex-methods}
%

\iffalse
\Pedro{GD again ???}
\GG{Done}
\fi

The taps clusterization process presented utilizes KNN, which is an unsupervised learning method that introduces an approximation error, creating a trade-off between complexity and performance (Fig.~\ref{fig:methodology-clustering-methods}b). This section presents the application of supervised learning through the GD algorithm to decrease this approximation error and achieve lower complexity.

% \begin{figure}[htbp]
%     \centering
%     \includegraphics[height=3.92cm]{optimized-centroids.pdf}
%     \caption{Clustered taps ($g_C[n]$) before (left) with 10 clusters and after gradient descent (right) with only 8 clusters, for 160km SMF, 32GBaud}
%     \label{fig:optimized-centroids}
% \end{figure}

\iffalse
\Pedro{instead of Hardware Implementation block? Not clear }
\GG{Done}

\Pedro{why before fiber?}
\GG{Done}

\fi

% A structure depicted in Fig.~\ref{fig:cluster-fpga} was employed for the training process. For the training, the \textit{"SOFTWARE IMPLEMENTATION"} scheme depicted in Fig.~\ref{fig:cluster-fpga}(left) was used to generate the clustered filter taps (centroids), by applying the KNN. A convolutional neural network (CNN) was implemented in software and its weights were initialized with the centroids obtained by KNN, the remaining training used as features (input data) the pre-summed values $x_S[n]$ calculated by Eq.~(\ref{clustered-summation}) using the samples obtained from the numerical simulation of SSMF, as explained previously, and as the labels (desired prediction) the samples before optical fiber propagation simulation. \J{a long and unclear sentence, rewrite}

To optimize the process, we utilized the clustering scheme shown in Fig.~\ref{fig:cluster-fpga}(left), identical to the one used for TDCE KNN, to produce clustered filter taps (centroids). We subsequently developed a convolutional neural network (CNN) in software, initializing it with these centroids. The CNN was trained using pre-summed values $x_S[n]$, calculated by Eq.~(\ref{clustered-summation}) from samples that simulated propagation through SSMF, as features (input data). The labels (desired predictions) were derived from the samples prior to optical fiber propagation simulation.

\iffalse
\Pedro{how many total epochs ? add that this early stop uses the remaining data from the training to validate}
\GG{Check: Total epochs added. We are not doing this validation.}

\Pedro{why just $2^{16}$ }
\GG{Check if it is ok}

\Pedro{refer to the true number of reduction by fine-tuning with gd}
\GG{Done}
\fi

The total experimental dataset used consisted of $2^{18}$ symbols for the training dataset and $2^{18}$  independently generated symbols for the testing phase, each with different random seeds. Each training epoch utilized $2^{16}$ randomly selected samples from the training dataset, ADAM optimizer with a learning rate of $10^{-5}$, 400 training epochs. A training ``stop condition'' of 50 epochs with no BER improvement was also implemented. The testing phase utilized the first $2^{16}$ samples from the testing dataset, as this quantity was enough for the model to continuously learn until the early stop condition was reached.

An interesting finding of this strategy is that the centroids optimized by supervised learning became outside of the circle as an effect of correcting the approximation error introduced by the clustering process, as shown in Fig.~\ref{fig:methodology-clustering-methods}(f).
As a result, this fine-tuning reduced the number of clusters needed to reach the pre-FEC threshold considered in this paper, leading to a lower complexity filter with complexity savings of 33.3\%, 20\% and 20\% for 1, 2, and 4 spans, respectively, in comparison to TDCE KNN and, 30.3\%, 14.3\%, 22.4\% in comparison to FDE, as illustrated in Fig.~\ref{fig:methodology-clustering-methods}(g).

\subsection{FDE. Methodology}\label{subsec:methodology-fde}
We aim to benchmark our solution against the simplest FDE available to ensure a fair comparison. We assessed the performance versus filter size as shown in Fig.~\ref{fig:methodology-clustering-methods}(a) for the time-domain FIR filtering, identifying the minimum filter size necessary to satisfy the condition \(BER<3.8\times10^{-3}\). Then, we determined the FFT size ($N_{FFT}$) that minimizes the complexity of the FDE, $C_{FFT}$:
\begin{equation}
\label{fft-complexity}
C\textsubscript{FFT} = N_{FFT} \, \frac{8 \beta \log_2(N_{FFT})+4}{N_{FFT}-M+1},
\end{equation}
where \(N_{FFT}\) represents the FFT size (block size), \(M\) denotes the filter size, and \(\beta\) is a constant equal to \(1/2\) when using the Radix-2 architecture with FFT sizes that are powers of 2, or \(\beta = 3/8\) when using the Radix-4 architecture with FFT sizes that are powers of 4. We utilized the overlap-save method\cite{xu2011frequency} to segment the input signal into blocks before applying the FDE. Therefore, the denominator of Eq.~(\ref{fft-complexity}) represents the total number of samples not corrupted by the overlap-save method. This optimization process of finding the optimization process is illustrated in the plot of Fig.~\ref{fig:methodology-clustering-methods}(c).

\iffalse
\Pedro{Geraldo to be clear you always need to follow the following logic - what we did, why we did it- what we obtained - the importance of the result}
\GG{Check: didn't understand}
\fi

By using the methodology described in this section and in the previous ones, the filter parameters in our approach were defined as described in Table~\ref{table:parameters-details}.
\begin{table}[h]
\centering
\setlength{\tabcolsep}{4pt} % Adjust column separation for a more compact table
\renewcommand{\arraystretch}{1.2} % Adjust row separation for readability
\resizebox{0.47\textwidth}{!}{
\begin{tabular}{ccccccc}%{|c|c|c|c|c|c|c|c|c|c|}
\hline
\textbf{ } & \textbf{N} & \textbf{M\textsubscript{TDCE}} & \textbf{M\textsubscript{FDE}} & \textbf{N\textsubscript{C}(KNN)} & \textbf{N\textsubscript{C}(GD)} & \textbf{N\textsubscript{FFT}}\\
\hline
1 span & 45 & 31 & 29 & 9 & 6 & 256\\
\hline
2 spans & 89 & 53 & 55 & 10 & 8 & 256\\
\hline
4 spans & 177 & 97 & 103 & 10 & 8 & 1024\\
\hline
8 spans & 353 & 189 & 201 & 12 & 12 & 4096\\
\hline
\end{tabular}}
\caption{Filter parameters details; see the body text for explanations.}
\label{table:parameters-details}
\end{table}
In that Table $N$ is the maximum filter size as defined by Eq.~(\ref{taps-bound}), $M_{TDCE}$ is the filter size to achieve $BER<1\times10^{-3}$ before clustering for both TDCE variations, $M_{FDE}$ is the filter size for FDE, $N_C(KNN)$ and $N_C(GD)$ are the number of complex clusters for the TDCE KNN and TDCE GD respectively and $N_{FFT}$ is the FFT size used for FDE.

%\J{Higher than what?}
%\J{the difference between what and what?} 

Although the pre-FEC threshold used to define the filter size for FDE was higher than the one for TDCE, it required slightly more taps ($M_{FDE}$) than the TDCE ($M_{TDCE}$) in some cases, as shown in Table~\ref{table:parameters-details}. As the difference in the number of filter taps was 6.3\% in the worst case (8 spans), we considered it negligible.

Having established the optimal number of clusters and FFT sizes, we proceed to calculate the complexity; the respective dependencies are given in Fig.~\ref{fig:methodology-clustering-methods}(g) for the FDE and two TDCE variants. Notably, employing a time-domain approach allows us to attain a complexity level comparable to that of the FDE as measured in real multiplications per recovered symbol.

\section{FPGA Implementation}\label{Sec:fpga}
%

\iffalse
\Pedro{is this right?, I would say: Since the accumulation of taps at a specific angle happens sequentially but not in a consecutive manner, it is crucial to route the samples correctly to the appropriate pre-summed array memory for proper accumulation.}
\GG{Done}
\fi

\subsection{Simplified Convolution}
As the index of the filter taps overlapped at a specific angle is not consecutive, it is crucial to route the input samples correctly to the appropriate pre-summed array memory \(x\textsubscript{S}[k]\) for the accumulation expressed in Eq.~(\ref{clustered-summation}). This routing presents a significant challenge similar to the routing challenge mentioned in Ref.~\cite{martins2016distributive} as one of the further optimizations required in that approach. Further on, we explain the implementation of this routing block in detail.

\begin{figure}[htbp]
    %\centering
    % \hspace{-5.5mm}
    \includegraphics[width=0.48\textwidth]{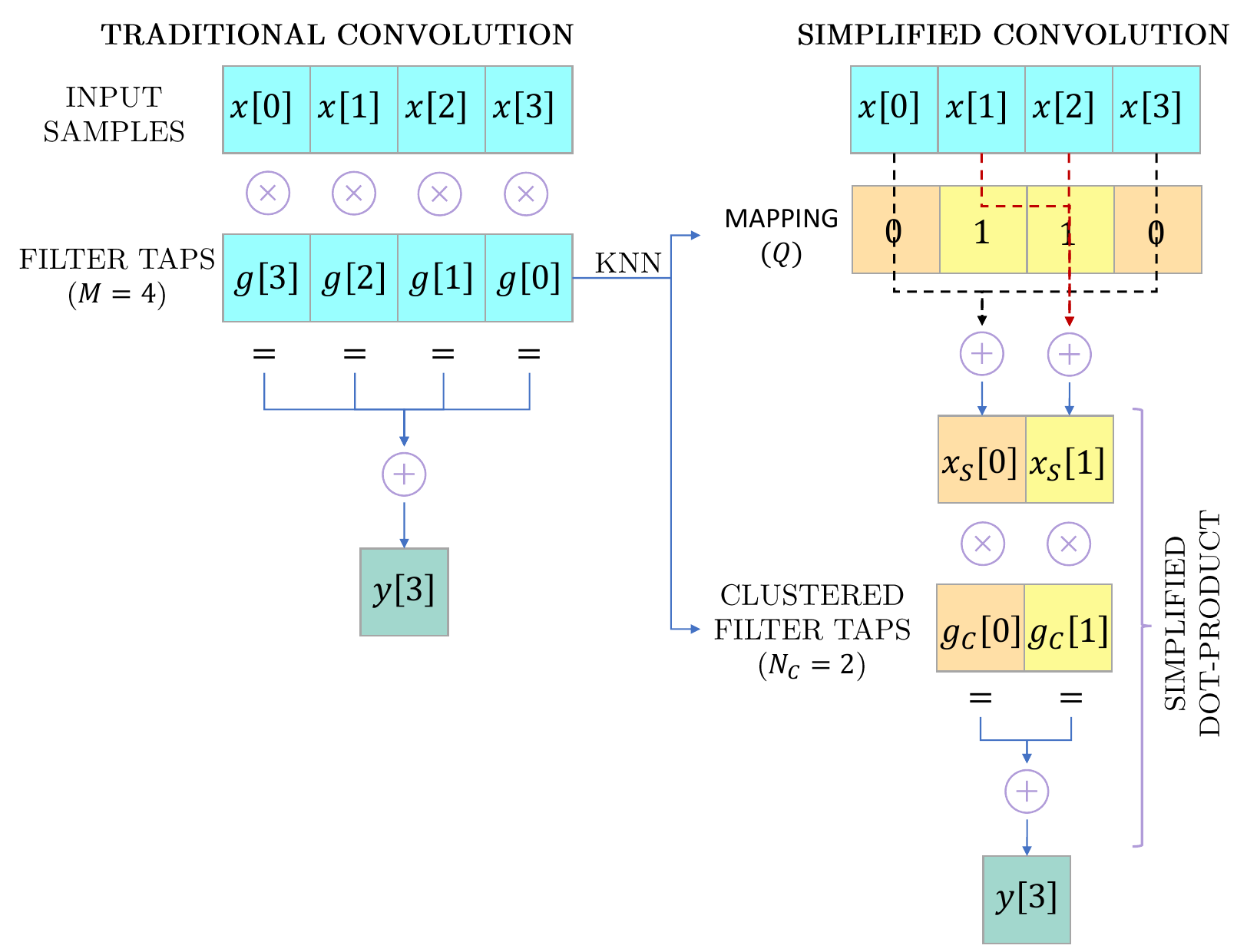}
    \caption{Conventional FIR filtering (left) compared to clustered filtering (right). After the clustering process (KNN), some filter taps acquire the same value. As there are $N_C$ distinct values, a sample mapping array ($Q$) can express to what cluster each sample belongs.}
    \label{fig:convolution-fpga-explanation}
\end{figure}

%\J{for which purpose?}

% The routing of input samples in our implementation is based on the ideas presented in Ref.~\cite{samragh2017customizing}, in which the equivalent idea to our $Q_j$ sets used in Eq.~(\ref{clustered-summation}), is a single set that acts like an encoding dictionary. This idea was used in our approach as well, and Fig.~\ref{fig:convolution-fpga-explanation} illustrates that now, only one set called $Q$ is used for the routing purpose instead of the sets $Q_j$, which allow us to read the samples sequentially and at the same time read from the single set $Q$ to which position of $x_S$ this sample must accumulated.

The sample routing of our implementation is inspired by the ideas presented in Ref.~\cite{samragh2017customizing}. Unlike the multiple $Q_j$ sets described in Eq.~(\ref{clustered-summation}), we use a single set, $Q$, as shown in Fig.~\ref{fig:convolution-fpga-explanation}. This set $Q$ contains pointers for each filter tap position, indicating where each input sample should be accumulated in $x_S$. This method enables sequential input sample reading and simultaneous routing using $Q$.

\iffalse

\Pedro{need to rewrite so the reader can understand better this entire paragraph }
\GG{Check if understandable now}

\fi

% Considering the graphical interpretation of convolution as the dot-product of the reversed sliding filter ($N$ taps) over the input samples, a hypothetical example of output calculation is given for $y[3]$ in Fig.~\ref{fig:convolution-fpga-explanation}(left). The clustering process using KNN, as explained in previous sections and illustrated in Fig.~\ref{fig:convolution-fpga-explanation} (between left and right sides), generates the centroids that will be used as the clustered filter taps ($N_C$ clusters) and the sample mapping ($Q$), which can be implemented as sequential memory and indicates to what pre-summed set \(x\textsubscript{S}\) each input sample must be accumulated. In this way, the routing can be done, and the sets \(x\textsubscript{S}\) can pass to the simplified dot-product stage illustrated in Fig.~\ref{fig:convolution-fpga-explanation}(right), which consists of a simplified dot-product between \(x\textsubscript{S}\) and the clustered filter taps \(g\textsubscript{C}\) harnessing the distributive property of multiplication. 

Fig.~\ref{fig:convolution-fpga-explanation}(left) illustrates a hypothetical output calculation for $y[3]$, treating convolution as the dot-product of a reversed sliding filter with $M$ taps over input samples. As detailed earlier and shown in Fig.~\ref{fig:convolution-fpga-explanation}, the KNN-based clustering process generates centroids for clustered filter taps ($N_C$ clusters) and sample mapping ($Q$). This mapping, implemented as sequential memory, directs each input sample to a pre-summed set \(x\textsubscript{S}\). This routing allows a simplified dot-product between \(x\textsubscript{S}\) and \(g\textsubscript{C}\), as depicted in Fig.~\ref{fig:convolution-fpga-explanation}(right), leveraging the distributive property of multiplication.

\begin{figure*}[h!]
    \centering
    % \hspace{-5mm}
    \includegraphics[width=1.0\linewidth]{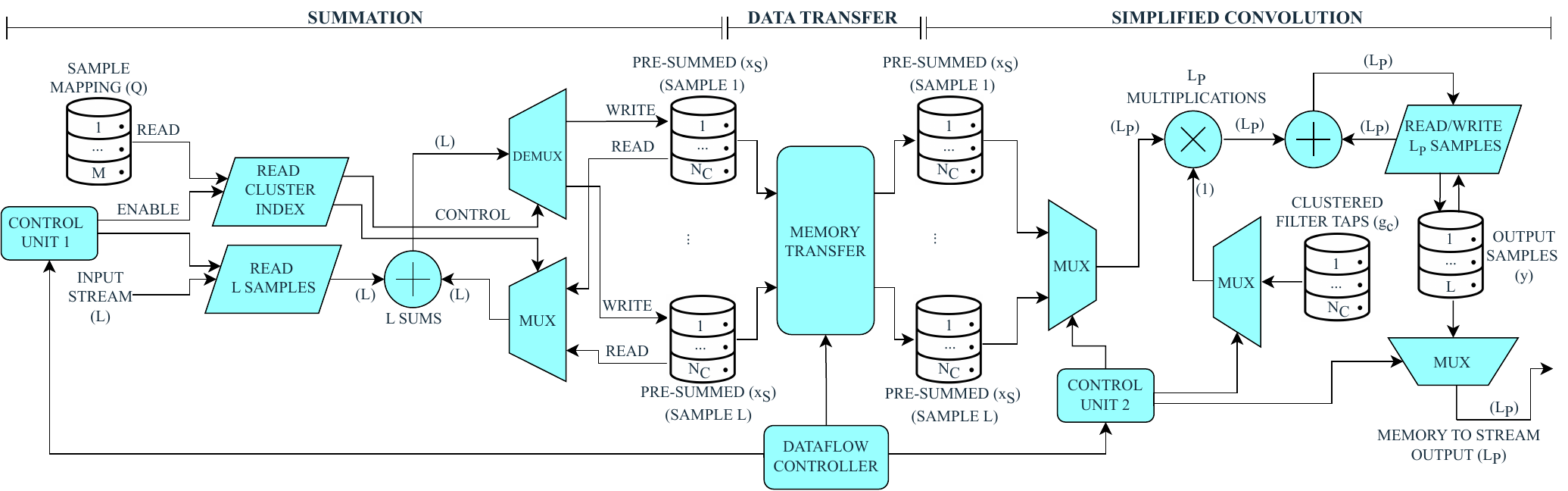}
    \put(-510,142){\textcolor{red}{\textbf{(a)}}}
    \put(-255,142){\textcolor{red}{\textbf{(b)}}}
    \put(-175,142){\textcolor{red}{\textbf{(c)}}}
    \caption{(a) Control Unit 1, enabled by the general Dataflow Controller, reads \textit{L} samples from input and the associated cluster index. It includes a memory for storing the sample mapping set (\textit{Q}) and \textit{L} memory banks with \textit{N\textsubscript{C}} positions to store the pre-summed values \(x\textsubscript{S}\). The cluster index controls the MUX to read and accumulate the new input values with the previously stored ones. A DEMUX operation updates the memory banks with the summed values.
    (b) The Memory Transfer block quickly transfers \(x\textsubscript{S}\) to another memory bank, freeing the summation memories for processing the next samples.
    (c) Control Unit 2, enabled by the Dataflow Controller, reads pre-summed values with the same cluster index from different memory banks in blocks of \textit{L\textsubscript{P}} samples. These values are directed to the multiplier via a MUX operation, multiplied by the respective cluster value, and accumulated. The output bank memory is updated until all \textit{L\textsubscript{P}} samples are ready, then outputted as a parallel stream.
}
    \label{fig:fpga-diagram}
\end{figure*}

\subsection{Parallelization}\label{subsec:parallelization}
\iffalse
\Pedro{Please highlight here and in the abstract introduction and conclusion that this is one of the novelties of the paper as well. Great insights here =)}
\GG{Done}
\fi

The parallelization of TDCE is a novel aspect of our research. To explain this process, we expand on the simplified convolution shown in Fig.~\ref{fig:convolution-fpga-explanation}. As the simplified dot-product is simple and has a straightforward implementation, we focus on the summation process that involves issues like routing and parallel memory access. The pseudo-code in Algorithm~\ref{alg:summation} illustrates the steps to obtain $x_S$, similar to the method in Ref.~\cite{samragh2017customizing}.

\begin{algorithm}
\caption{Pseudo-code for summation}
\label{alg:summation}
\begin{algorithmic}[1]
\STATE \textbf{input:} $x, Q$
\STATE \textbf{output:} $x_S$
\STATE $x_S \leftarrow 0$
\FOR{$i=0...(M-1)$}
    %\IF{condition is met}
    \STATE $w = Q[i]$ //reading the cluster index
    \STATE $x_S[w]$ += $x[i]$ //accumulation
    %\ELSE
    %\ENDIF
\ENDFOR
\STATE \textbf{Return:} $x_S$
\end{algorithmic}
\end{algorithm}

\begin{figure}[H]
    %\centering
    \includegraphics[width=0.48\textwidth]{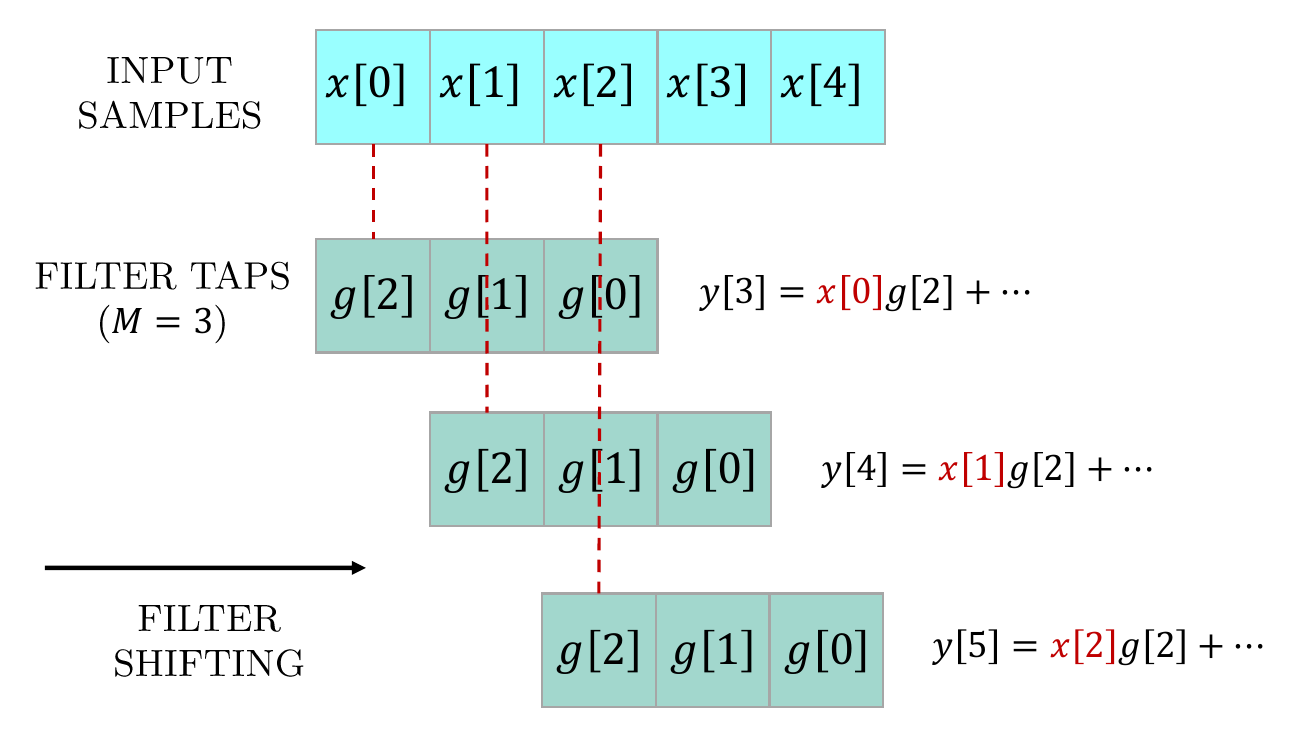}
    % \caption{Parallelization explanation. Some steps of the linear convolution are depicted for hypothetical output samples (y[3],y[4], and y[5]). We highlight that sequential input samples are multiplied by the same filter tap for each output sample, which means that they are associated with the same cluster index. This fact makes possible a high degree of parallelization with a sequential routing of values to produce \(x\textsubscript{S}[k]\). \J{It is not a good caption. The caption must clearly explain what we see in the figure - it is not here. But you give a lot of explanations -- the explanations must be moved into the main text, the caption must contain only the concise explanations for the object depicted in the fig.}}
    \caption{Property used for parallelization. Hypothetical steps to calculate the output samples ($y[3]$, $y[4]$, and $y[5]$) are shown. Red dashed lines emphasize how the same filter tap, $g[2]$, multiplies successive input samples across these outputs. For example, for $y[3]$, $g[2]$ multiplies $x[0]$, while for the consecutive output $y[4]$, $g[2]$ multiplies the subsequent input $x[1]$.}

    \label{fig:parallelization-fpga}
\end{figure}

A possible strategy to boost computation speed is to fully unroll the loop in Algorithm~\ref{alg:summation} and execute all iterations in parallel. This requires reading $M$ positions of $Q$ and accessing $M$ input sample positions simultaneously for accurate accumulation into \(x\textsubscript{S}[k]\). While this increases computational speed, it significantly increases hardware resource demands due to parallel random memory access and numerous sum operations. To efficiently use FPGA resources, we compute $L$ sequential output samples simultaneously rather than parallelizing each output sample. This optimization leverages the interpretation of time-domain convolution as a series of dot-products, detailed in Fig.~\ref{fig:parallelization-fpga}.

Fig.~\ref{fig:parallelization-fpga} illustrates convolution as a series of dot-products between sliding reversed filter taps and input samples, highlighting that the same filter tap multiplies consecutive input samples for successive outputs. Each filter tap in the original filter belongs to a cluster in the clustered filter, and each input sample must be routed to its corresponding clustered filter tap, as shown in Fig.~\ref{fig:convolution-fpga-explanation}. Since $L$ consecutive input samples are multiplied by the same filter tap across $L$ consecutive outputs, all $L$ input samples must be mapped to the same cluster. To utilize this property, we developed Algorithm~\ref{alg:summation-L}.

\begin{algorithm}
\caption{Pseudo-code for parallelized summation}
\label{alg:summation-L}
\begin{algorithmic}[1]
\STATE \textbf{input:} $x, Q$
\STATE \textbf{output:} $x_S$
\STATE $x_S \leftarrow 0$
\FOR{$i=0...(M-1)$}
    %\IF{condition is met}
    \STATE $w = Q[i]$ //reading the cluster index
    \FOR{$j=0...(L-1)$}
    %\IF{condition is met}
    \STATE $x_S[j][w]$+=$x[i+j]$ //accumulation

    \ENDFOR
\ENDFOR
\STATE \textbf{Return:} $x_S$
\end{algorithmic}
\end{algorithm}

Algorithm~\ref{alg:summation-L} leverages the property from Fig.~\ref{fig:parallelization-fpga}, where the cluster index is read from $Q$ once and used for all the innermost loop iterations. This property enables sequential reading of the input sample array $x$. Since each output sample requires its own $x_S$ set, $x_S$ is now a bidimensional array in the algorithm. Additionally, the innermost loop is fully parallelizable, reducing the total number of iterations for the complete algorithm to $M$. This significantly enhances throughput, making the total time period to perform the algorithm independent of $L$.

%\J{which architecture?}

\subsection{TDCE Hardware Implementation Architecture}
The hardware implementation of the TDCE is summarized in Fig.~\ref{fig:fpga-diagram}. To achieve high throughput, the architecture leverages the fact that the structure producing the pre-summed array \(x\textsubscript{S}[k]\) is idle during the simplified dot-product stage, as shown in Fig.~\ref{fig:convolution-fpga-explanation}(right). This allows this structure to start processing the next samples while the simplified convolution of the previous ones is still ongoing. The Dataflow Controller block manages this process by controlling the Control Unit blocks for summation and simplified convolution.

%\J{in the end?}
In the summation structure, the sample mapping ($Q$) is stored in a register and read sequentially for each filter tap ($M$). Control Unit 1, enabled by the Dataflow Controller, enables the parallel reading of $L$ samples and one mapping simultaneously, $M$ times. This approach minimizes random access, as explained in Section~\ref{subsec:parallelization}. By pipelining this structure, we can organize the loop tasks—reading input, reading cluster index, reading memory banks, accumulation, and updating memory banks—to start a new iteration loop every clock cycle.

% \J{not a good sentence after "releases" - it is unclear to which subject the verb "releases" belongs}

Controlled by the Dataflow Controller, the Memory Transfer block performs a bulk transfer of data from the summation memory banks (left) to secondary memory banks (right). This frees up the summation memory banks to process the next $L$ samples. Consequently, the structure can process new input samples before the simplified dot-product is finished.

% \J{out of context sentence - the paragraph started from the reference to Dot Prod - and then all of a sudden we have "However" - in which process the num of mult is less than additions? explain more carefully}

%\J{Both? Unclear}
%\J{YOu used the capitalisation like "Simplified Dot-Product" - be consistent all over the text}

After the Summation process is completed, the Simplified Dot-Product is executed. The number of multiplications in the Simplified Dot-Product is significantly less than the number of additions in the Summation, allowing multiplications to be performed at a lower rate, thus saving multipliers. For $L$ input samples, we define the number of parallel samples $L_P$ to be multiplied such that $L$ is a multiple of $L_P$. This strategy ensures that the Simplified Dot-Product takes a similar amount of time as the summation process, enabling both operations to overlap. Control Unit 2 enables the reading of blocks of size $L_P$ to be multiplied by the same clustered filter tap. The resulting $L_P$ samples are accumulated in an output memory, which, at the end of the process, contains the $L$ output equalized samples. This multiply-and-accumulate (MAC) process is done for each cluster, resulting in $N_C$ iterations. Each iteration consists of $L/L_P$ internal iterations to perform MAC operations for all $L$ samples. Due to the overlapped execution, the system's throughput (\textit{TH}) is given by:
\begin{equation}
\label{eq:throughput}
TH = \frac{L}{\alpha M},
\end{equation}
where \(\alpha\) is an empirical constant defined by the dataflow overlapping, and \textit{M} is the filter size.

\subsection{FDE Hardware Implementation Architecture}

%\J{Split, avoid tautology: "used" is used 2 times consequently}

\begin{figure}[htb]
    %\centering
    \hspace{-3mm}
    \includegraphics[width=0.5\textwidth]{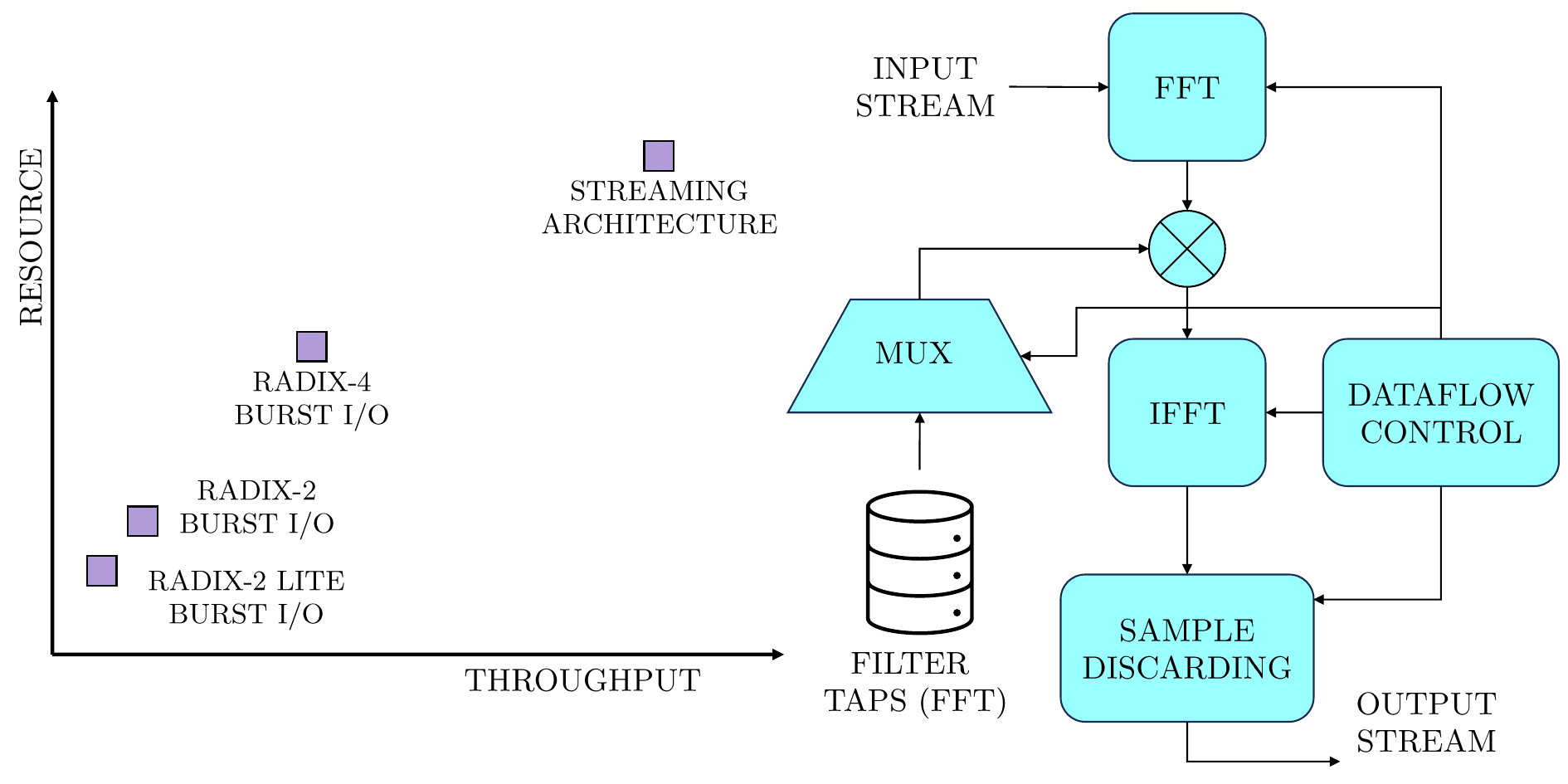}
    \caption{FDE implementation details. Left: throughput versus resources usage trade-off among different FFT architectures from Xilinx FFT IP module~\cite{xilinx2024fft}. Right: FDE block diagram for FPGA implementation with dataflow control for tasks overlapping and improved throughput.}
    \label{fig:fde-equalizer}
\end{figure}

The FDE is used for comparison in this work because it is the common state-of-the-art reference in previous studies~\cite{martins2016distributive,felipe2020chirp,ji2023hardware}. It offers the lowest complexity (calculated by Eq.~(\ref{fft-complexity})) compared to time-domain FIR filters~\cite{spinnler2010equalizer}, a higher degree of parallelization and stability compared to IIR filters~\cite{spinnler2010equalizer}, and is preferred for ASIC implementations~\cite{sun2020800g,fougstedt2020asic}. The FDE is based on the FFT algorithm, which includes a forward FFT to transform the input signal block, multiplication in the frequency domain (FD) between the FD representations of the input signal and transfer function, followed by an IFFT operation, and discarding of corrupted samples by the circular convolution (Fig.~\ref{fig:fde-equalizer} right)~\cite{xu2010chromatic}. To implement FDE, a method for block division and subsequent filtering of the signal is required, with Overlap-Add (OA) and Overlap-Save (OS) being well-known options~\cite{xu2011frequency}. In this work, we use OS for our FDE implementation. However, hardware resources for block division were not considered, as we aim to compare only the mathematical operations of the filters.

\iffalse
 \Pedro{ref for this statement}
 \GG{check: I don't think it is possible}
\fi
%\J{a solution of TDCE? Is it a standard term? Maybe "a variant" or "a realization"?}

To compare our solution with the state-of-the-art FDE, we used the FFT HLS Library~\cite{xilinx2024fft} developed and optimized by \AMD{} for the FFT/IFFT blocks. This ensures an unbiased and fair comparison since it was developed independently of our research. This FFT module offers different architecture options, allowing users to choose the suitable trade-off between hardware resources and throughput (Fig.~\ref{fig:fde-equalizer} left). In this work, we used the Radix 4 Burst/IO architecture for medium output throughput, as the high throughput realization of TDCE will be addressed in future studies.

%\J{What??}
%\J{By doing what? Mitigation? Dot-product? Outputting? Subsequent latency?}

The Radix 4 Burst/IO architecture is not pipelined and has a latency before outputting data for each transformation cycle. To mitigate this and achieve higher throughput for FDE, we applied a dataflow HLS directive to implement the structure shown in Fig.~\ref{fig:fde-equalizer}(right). This allows the FFT block's data output to overlap with the FD multiplication and the IFFT's subsequent latency. Using dataflow, the throughput for FDE can be estimated by:

\begin{equation}
\label{eq:throughput-fde}
TH = \frac{N_\textsubscript{FFT}-M+1}{\delta\Delta\textsubscript{FFT}},
\end{equation}
in which \(\Delta\textsubscript{FFT}\) is the latency for the FFT size \(N_\textsubscript{FFT}\) used and \(\delta\) is an empirical constant determined by the dataflow structure.

% \J{Two contradicting specifications in one sentence: for simplicity... for fairness reasons... So, two rabbits in one shot? Clumsy and unclear whether you wanted simplicity or fairness}

As the IP block from \AMD{} does not allow a controllable parallelization, for simplicity, the throughput of TDCE was adjusted accordingly to \textit{match the same FDE throughput} by adjusting the $L$ parameter in Eq.~(\ref{eq:throughput}).

\iffalse
\Pedro{highlight in bold that all throughputs are the same for fairness reasons}
\GG{Done}
\fi

\subsection{Implementation Parameters and Methodology}

In all subsequent analyses, we implemented both filters on a VCK190 FPGA by simulations in the \AMD{} Vitis IDE at a clock frequency of 250 MHz. Both filters applied identical quantization to the taps and signals to ensure a fair comparison. A total of 16 bits, with 5 for integer bits, were used for the TDCE, and a total of 16 bits, with 1 integer bit, were used for the FDE to meet the threshold. All the filters were implemented aiming a BER$<3.8\times10^{-3}$, and the achieved BER values for the parameters used are described in Table~\ref{table:performance-details}.
\begin{table}[htb]
\centering
\setlength{\tabcolsep}{4pt} % Adjust column separation for a more compact table
\renewcommand{\arraystretch}{1.2} % Adjust row separation for readability
\resizebox{0.40\textwidth}{!}{
\begin{tabular}{cccc}%{|c|c|c|c|c|c|c|c|c|c|}
\hline
\textbf{ } & \textbf{BER\textsubscript{GD}} & \textbf{BER\textsubscript{KNN}} & \textbf{BER\textsubscript{FDE}} \\
\hline
1 span & $2.45 \times 10^{-4}$ & $1.35 \times 10^{-3}$ & $1.69 \times 10^{-3}$ \\
\hline
2 spans & $3.32 \times 10^{-3}$ & $2.28 \times 10^{-3}$ & $9.31 \times 10^{-4}$ \\
\hline
4 spans & $2.79 \times 10^{-3}$ & $2.30 \times 10^{-3}$ & $8.87 \times 10^{-4}$ \\
\hline
8 spans & $3.39 \times 10^{-3}$ & $3.21 \times 10^{-3}$ & $2.01 \times 10^{-3}$ \\
\hline
\end{tabular}}
\caption{Performance comparison.}
\label{table:performance-details}
\end{table}

Additionally, all the TDCE implementations were designed to achieve the same throughput as FDE with a maximum error difference of less than 7\%, balancing the trade-off between digital filter throughput and hardware resources. The parameters of the parallelization of TDCE approaches and the achieved throughput are summarized in Table~\ref{table:throughput-details}.

\begin{table}[H]
\centering

\setlength{\tabcolsep}{4pt} % Adjust column separation for a more compact table
\renewcommand{\arraystretch}{1.2} % Adjust row separation for readability
\resizebox{0.5\textwidth}{!}{
\begin{tabular}{cccccccc}
\hline
\textbf{} & \textbf{L\textsubscript{GD}} & \textbf{L\textsubscript{KNN}} & \textbf{L\textsubscript{P-GD}} & \textbf{L\textsubscript{P-KNN}} & \textbf{TH\textsubscript{GD}} & \textbf{TH\textsubscript{TDCE}} & \textbf{TH\textsubscript{FDE}} \\
\hline
1 span & 8 & 10 & 2 & 2 & $81.6$ & $83.3$ & $80.5$ \\
\hline
2 spans & 10 & 12 & 2 & 2 & $66.7$ & $75.9$ & $71.3$ \\
\hline
4 spans & 18 & 20 & 2 & 2 & $75.6$ & $81.3$ & $79.5$ \\
\hline
8 spans & 36 & 36 & 2 & 2 & $76.9$ & $76.9$ & $78.6$ \\
\hline
\end{tabular}}
\caption{Performance Metrics. All throughputs are expressed in Mb/s.}
\label{table:throughput-details}
\end{table}
%\J{Delete the astericses}

% \J{The power is associated? To which word does your "which" refer?} 
The equalizers were implemented using C++ in Vitis HLS 2023.2, which translates the code to HDL and generates RTL. The RTL was then implemented in Vivado 2023.2 using the default routing strategy. Energy consumption was assessed with the Power Estimator tool in Vivado. Power consumption in digital circuits comprises static and dynamic power. We focused on the dynamic power of the filters, excluding static power, which is associated with all inner circuits regardless of their usage in FPGA implementations. Additionally, dynamic power is the dominant factor in power consumption for ASIC implementations~\cite{fougstedt2020asic}.

\subsection{Memory Impact}\label{subsec:memory}
Memory requirements and their hardware implementation are often not discussed in the evaluation of proposed algorithms for CDC~\cite{felipe2020chirp,martins2016distributive,wu2022low}. Here, we demonstrate that this is an important issue to be considered that can impact the hardware implementation results noticeably. 

%\J{Unclear - which details? Looks like a non-scientific jargon again}
% \J{Too many paranthesis, too complicated a sentence - rewrite clearly}
% \J{What is 81? Do not use digits without explanations - name explicitly}
% \J{A strange sentence - correct and rewrite}

To analyze the impact of memory implementation, TDCE KNN and FDE were implemented on FPGA, following the steps in Section~\ref{sec:methodology} for a fiber length of 4 spans (320~km). Results are shown in Fig.~\ref{fig:memory-impact}. The implementation used two types of memory for the bank memories of \(x\textsubscript{S}[k]\) (Fig.~\ref{fig:fpga-diagram}) in TDCE and inside the FFT/IFFT blocks of FDE (Fig.~\ref{fig:fde-equalizer} right). Table~\ref{table:parameters-details} shows that for TDCE, \(L=20\) was used to achieve 81.3Mb/s, while for FDE, \(N\textsubscript{FFT}=1024\) was used to achieve 79.5Mb/s.

\begin{figure*}[ht!]
    \centering
    \begin{minipage}[t]{0.22\linewidth}
    \centering
    \vspace{-5mm} % Adjust spacing as needed
        \includegraphics[height=5cm]{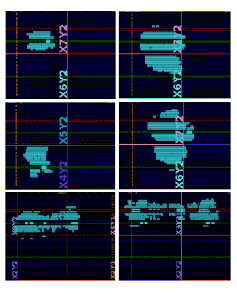}
        \put(-120,99){\textcolor{red}{\rotatebox{90}{\textbf{\fontsize{7}{16}\selectfont TDCE GD}}}}
        \put(-120,52){\textcolor{red}{\rotatebox{90}{\textbf{\fontsize{7}{16}\selectfont TDCE KNN}}}}
        \put(-120,20){\textcolor{red}{\rotatebox{90}{\textbf{\fontsize{7}{16}\selectfont FDE}}}}
        \put(-95,0){\textcolor{red}{\rotatebox{0}{\textbf{\fontsize{7}{16}\selectfont 1SPAN}}}}
        \put(-45,0){\textcolor{red}{\rotatebox{0}{\textbf{\fontsize{7}{16}\selectfont 8SPANS}}}}
        \put(-132,130){\textcolor{red}{\textbf{(a)}}}
    \end{minipage}
    \hspace{0.02\linewidth} % Adjust spacing as needed
    \begin{minipage}[t]{0.22\linewidth}
    \vspace{-3.5mm}
    \centering
        \begin{tikzpicture}[scale=0.56]
    \hspace{-7mm}
            \begin{axis}[
                ybar,
                symbolic x coords={LUT, FF, BRAM, DSP},
                xtick=data,
                ylabel={\textbf{LUT/FF(Qty/1000)}},
                ylabel style={yshift=-5pt,xshift=-5pt,font=\large},
                bar width=10pt,
                xtick align=inside,
                enlarge x limits=0.2,
                legend style={at={(0.51,0.95)}, anchor=north,legend columns=-1},
                % nodes near coords,
                % nodes near coords align={vertical},
                ymin=0,
                ymax=6500,
                axis y line*=left,
                axis x line*=none,  %because the other axis will add this
                yticklabel style={
                    /pgf/number format/.style={
                        fixed,
                        fixed zerofill,
                        precision=0,
                        scientific auto,
                        tick label style={/pgf/number format/1000 sep=,}
                    },font=\large
                },
                ytick={0, 1000, 2000, 3000, 4000, 5000, 6000},
                yticklabels={0, 1, 2, 3, 4, 5, 6},
                xticklabels={}, %because the other axis will add this
                ymajorgrids=true,
                grid style={dashed},
                height=9cm, % Set the height of the plot
                width=8cm,
                legend image code/.code={%
                    \draw [#1] (0.1cm,-0.1cm) rectangle (0.4cm,0.4cm); % Draw a square
                }
            ]
            % Data for TDCE GD (left y-axis)
            \addplot [
                    ybar,
                    fill=blue!30!white,
                    draw=blue!80!white, line width=0.5pt, %
                    postaction={
                        pattern=north east lines,
                        pattern color=blue!60!white
                        % pattern thickness=3pt, % Adjust thickness
                        % pattern width=6pt % Adjust pattern density
                    }
            ] coordinates {(LUT, 3085) (FF, 2918) (BRAM, 0) (DSP, 0)};
            \addlegendentry{TDCE GD}

            % Data for TDCE KNN (left y-axis)
            \addplot[
                    ybar,
                    fill=red!30!white,
                    draw=red!80!white, line width=0.5pt, %
                    postaction={
                        pattern=horizontal lines,
                        pattern color=red!60!white
                    }
            ]  coordinates {(LUT, 3415) (FF, 3192) (BRAM, 0) (DSP, 0)};
            \addlegendentry{TDCE KNN}
        
            % Data for FDE (left y-axis)
            \addplot[
            ybar,
            fill=green!30!white,
            draw=green!80!black, line width=0.5pt, %
            postaction={
            pattern=north west lines,
            pattern color=green!60!white
            }
            ]  coordinates {(LUT, 2800) (FF, 4755) (BRAM, 0) (DSP, 0)};
            \addlegendentry{FDE}
            \end{axis}
                            \node[text width=3cm] at (1.6,7.4) 
            {\textcolor{red}{\textbf{(b)}}};
            \begin{axis}[
                ybar,
                bar width=10pt,
                symbolic x coords={LUT, FF, BRAM, DSP},
                xtick=data,
                axis y line*=right,
                % axis x line=none,
                % nodes near coords,
                % nodes near coords align={vertical},
                xtick align=inside,
                ymin=0,
                ymax=19.5,
                ylabel={\textbf{BRAM/DSP}},
                ylabel style={yshift=10pt,xshift=-40pt,font=\large},
                yticklabel style={
                    /pgf/number format/.style={
                        fixed,
                        fixed zerofill,
                        precision=0,
                        scientific auto,
                        tick label style={/pgf/number format/1000 sep=,}
                    },font=\large
                },
                enlarge x limits=0.2,
                xticklabel style ={xshift=10pt, yshift=-4pt, font=\fontsize{13}{14}\selectfont,              rotate=45, % Rotate labels 45 degrees
                anchor=east % Align labels to the right
                },
                height=9cm, % Set the height of the plot
                width=8cm]
            % Data for TDCE GD (right y-axis)
            \addplot [
                    ybar,
                    fill=blue!30!white,
                    draw=blue!80!white, line width=0.5pt, %
                    postaction={
                        pattern=north east lines,
                        pattern color=blue!60!white
                    }
            ]  coordinates {(LUT, 0) (FF, 0) (BRAM, 0) (DSP, 4)};
            %\addlegendentry{TDCE GD}
            
            % Data for TDCE KNN (right y-axis)
            \addplot[
                    ybar,
                    fill=red!30!white,
                    draw=red!80!white, line width=0.5pt, %
                    postaction={
                        pattern=horizontal lines,
                        pattern color=red!60!white
                    }
            ]   coordinates {(LUT, 0) (FF, 0) (BRAM, 0) (DSP, 4)};
            %\addlegendentry{TDCE KNN}
            
            % Data for FDE (right y-axis)
            \addplot[
            ybar,
            fill=green!30!white,
            draw=green!80!black, line width=0.5pt, %
            postaction={
            pattern=north west lines,
            pattern color=green!60!white
            }
            ]  coordinates {(LUT, 0) (FF, 0) (BRAM, 8) (DSP, 14)};
            %\addlegendentry{FDE}

            \end{axis}

                \draw[thick, <->,red] (5.5,1.5) -- (5.5,5.3);
                \node[text width=1.5cm] at (4.8,4.3) 
            {\textcolor{red}{\footnotesize $\approx$ 71.4\%}};
            
        \end{tikzpicture}
    \end{minipage}
    \hspace{0.02\linewidth} % Adjust spacing as needed
    \begin{minipage}[t]{0.22\linewidth}
    % \vspace{-37mm}
        \vspace{-3.5mm}
        \centering
        \begin{tikzpicture}[scale=0.56]
        \hspace{-7.5mm}
            \begin{axis}[
                ybar,
                symbolic x coords={Clocks, Signals, Logic, BRAM, DSP},
                xtick=data,
                xticklabel style ={xshift=14pt, yshift=-4pt,font=\fontsize{12}{14}\selectfont,                rotate=45, % Rotate labels 45 degrees
                anchor=east % Align labels to the right
                },
                ylabel={\textbf{Power(mW)}},
                ylabel style={yshift=-6pt,xshift=0pt,font=\fontsize{13}{14}\selectfont},
                bar width=8pt,
                xtick align=inside,
                enlarge x limits=0.15,
                legend style={at={(0.51,0.95)}, 
                anchor=north,
                legend columns=-1},
                % legend image code/.code={\draw [fill=blue!30, draw=blue!80!black, line width=0.5pt, pattern=north east lines, pattern color=blue!60] (0,0) rectangle (0.8cm,0.8cm);}
                % },                
                % nodes near coords,
                % nodes near coords align={vertical},
                ymin=0,
                ymax=80,
                ytick={0, 10, 20, 30, 40, 50, 60, 70, 80},
                yticklabel style ={font=\fontsize{12}{14}\selectfont},
                ymajorgrids=true,
                grid style={dashed},
                height=9cm, % Set the height of the plot
                width=8.5cm,
                legend image code/.code={%
                    \draw [#1] (0.1cm,-0.1cm) rectangle (0.4cm,0.4cm); % Draw a square
                }
                % axis x line*=bottom,
                % axis y line*=left, % Ensure left y-axis line style is consistent
                % axis line style={-} % This makes the axis line style consistent
            ]
            % Data for TDCE GD
            \addplot [
                    ybar,
                    fill=blue!30!white,
                    draw=blue!80!white, line width=0.5pt, %
                    postaction={
                        pattern=north east lines,
                        pattern color=blue!60!white
                    }
            ]
            coordinates {(Clocks, 26) (Signals, 14) (Logic, 18) (BRAM, 0) (DSP, 15)};
            \addlegendentry{TDCE GD}

            % Data for TDCE KNN
            \addplot[
                    ybar,
                    fill=red!30!white,
                    draw=red!80!white, line width=0.5pt, %
                    postaction={
                        pattern=horizontal lines,
                        pattern color=red!60!white
                    }
            ] coordinates {(Clocks, 30) (Signals, 20) (Logic, 23) (BRAM, 0) (DSP, 16)};
            \addlegendentry{TDCE KNN}
        
            % Data for FDE
            \addplot[
            ybar,
            fill=green!30!white,
            draw=green!80!black, line width=0.5pt, %
            postaction={
            pattern=north west lines,
            pattern color=green!60!white
        }
        ] coordinates {(Clocks, 42) (Signals, 56) (Logic, 44) (BRAM, 27) (DSP, 57)};
            \addlegendentry{FDE}
            
            \end{axis}
            \node[text width=3cm] at (1.3,7.4) 
            {\textcolor{red}{\textbf{(c)}}};

            % \draw[red!80!pink,dashed] (0.0,1.3) -- (2,1.3);
            % \draw[red!80!pink,dashed] (0.0,4.4) -- (2,4.4);

        \end{tikzpicture}
    \end{minipage}
    \hspace{0.02\linewidth} % Adjust spacing as needed
    \begin{minipage}[c]{0.22\linewidth}    
    \vspace{-9mm}
    \begin{picture}(100,16)
    \hspace{-5mm}
        \put(-5,-145){
         \begin{tikzpicture}[scale=0.56]
         % \vspace{-100mm}
            \begin{axis} [
                xlabel={Number of Spans (80km)},
                xlabel style={yshift=-5pt,font=\fontsize{13}{14}\selectfont},
                ylabel={\textbf{(nJ)/bit}},
                grid=both,  
                ylabel near ticks,
                ylabel style={yshift=-15pt,xshift=0pt,font=\fontsize{13}{14}\selectfont},
                yticklabel style={
                    /pgf/number format/.style={
                        fixed,
                        fixed zerofill,
                        precision=0,
                        scientific auto,
                        tick label style={/pgf/number format/1000 sep=,}
                    },font=\large
                },
                xmin=1, xmax=8,
            	xtick={0, ..., 8},
            	ymin=0, ymax=4,
                legend style={legend pos=north east, 
                legend cell align=left,
                fill=white, 
                fill opacity=0.6, 
                draw opacity=1,
                text opacity=1,
                legend columns=-1},
                grid style={dashed},
                xticklabel style ={font=\fontsize{13}{14}\selectfont},
                height=9cm, % Set the height of the plot
                width=8.5cm]
                % ]

            \addplot[
                color=blue,
                mark=square*, % Change to desired shape
                mark size=3.5pt, % Increase the size of the marks
                very thick
            ] coordinates {
            (1, 0.56)(2, 0.80)(4, 0.98)(8, 1.14) %0.85,1.00,1.47,2.45
            };
            \addlegendentry{TDCE GD};
            
            \addplot[
                color=red,
                mark=*, % Change to desired shape
                mark size=3.5pt, % Increase the size of the marks
                very thick
            ]     coordinates {
            (1, 0.68)(2, 0.84)(4, 1.08)(8, 1.24) %0.85,1.00,1.47,2.45
            };
            \addlegendentry{TDCE KNN};
            
            \addplot[
                color=green,
                mark=triangle*, % Change to desired shape
                mark size=3.5pt, % Increase the size of the marks
                very thick
            ]   
            coordinates {
            (1, 2.34)(2, 2.55)(4, 2.78)(8, 3.15) %2.34,2.55,2.84,3.15
            };
            \addlegendentry{FDE};
            
                \end{axis}
                \node[text width=3cm] at (1.6,7.4) 
            {\textcolor{red}{\textbf{(d)}}};
            \draw[black!80!pink,dashed] (0.0,1.3) -- (2,1.3);
            \draw[black!80!pink,dashed] (0.0,4.4) -- (2,4.4);
            \draw[thick, <->,red] (1.8,1.3) -- (1.8,4.4);
                \node[text width=1.5cm] at (3.6,3.3) 
            {\textcolor{red}{\footnotesize $\approx$ 70.7\%}};
        \end{tikzpicture}    
        }
    \end{picture}
%     \begin{picture}(110,100)
%     \hspace{-1mm}
% \put(110,00){ \includegraphics[width=0.75\linewidth, height=0.27\linewidth]{chip.pdf}}
% \put(115,125){\textcolor{red}{\textbf{(c)}}}
% \put(156,125){\textcolor{red}{\textbf{2 spans}}}
% \put(266,125){\textcolor{red}{\textbf{4 spans}}}
% \put(376,125){\textcolor{red}{\textbf{8 spans}}}
% \put(453,108){\textcolor{red}{\rotatebox{270}{\textbf{TDCE}}}}
% \put(453,45){\textcolor{red}{\rotatebox{270}{\textbf{FDE}}}}
% \end{picture}
    \end{minipage}
    \caption{\footnotesize \textbf{(a)} Chip area implemented on FPGA, showing that TDCE has better area efficiency than FDE for low dispersion scenarios and similar area usage for higher dispersion. \textbf{(b)} Quantity of hardware resources used by all equalizers for 4 spans showing that TDCE uses 71,4\% fewer DSPs despite having bigger complexity, and the memory implementation of TDCE caused it to use fewer BRAMs but more LUTs than FDE. \textbf{(c)} Distribution of energy consumption per resource for all equalizers at 4 spans (640 km) showing that the energy efficiency of TDCE is better for all resource kinds. \textbf{(d)} Energy efficiency in nJ/bit depicting up to 70.7\% energy saving.}
    \label{fig:resources}
\end{figure*}
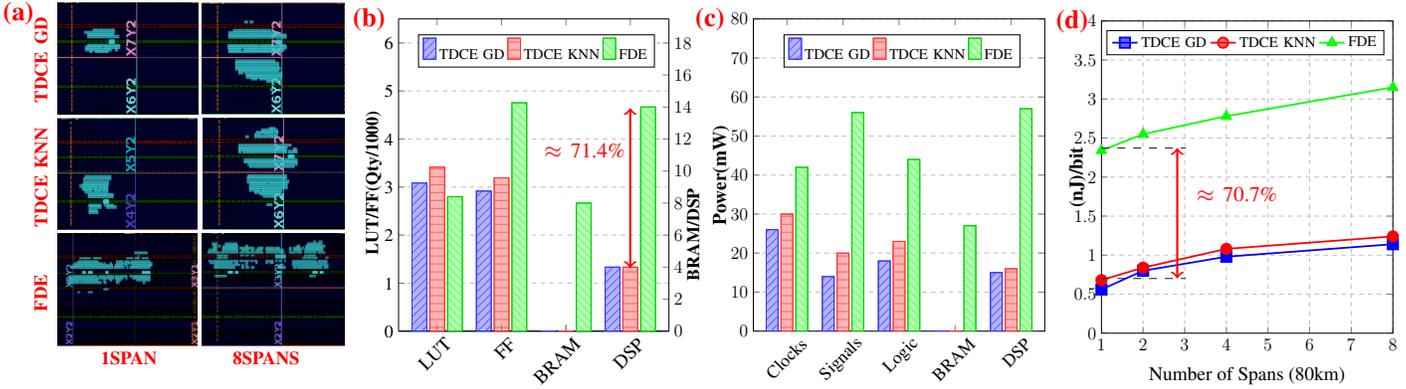

%\J{Less than what?}
%\J{Is raised by whom? A bead sentence - rewrite}
%\J{not affected by what?}

Fig.~\ref{fig:memory-impact}(a,b) illustrates the impact of using BRAMs for TDCE memory implementation. With \(L=20\) parallel samples processed, the total number of BRAMs is 40, as \(L\) memory banks are used before and after the Memory Transfer block (Fig.~\ref{fig:fpga-diagram}). This configuration reduces the quantities of LUTs and FFs compared to using LUTRAM, while the number of DSPs remains unaffected. However, dynamic power is higher when using BRAMs.

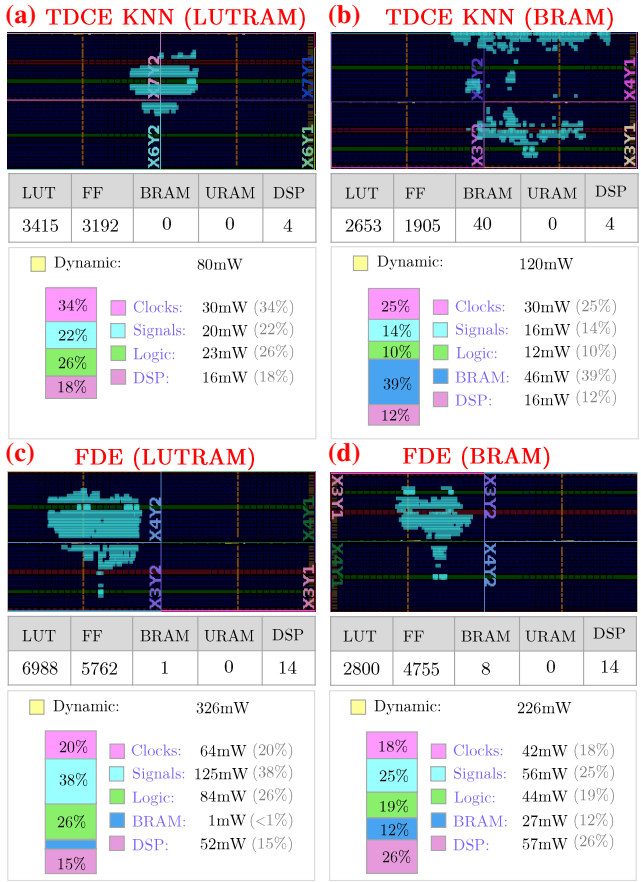
\begin{figure}[H]
    %\centering
    \includegraphics[width=0.48\textwidth]{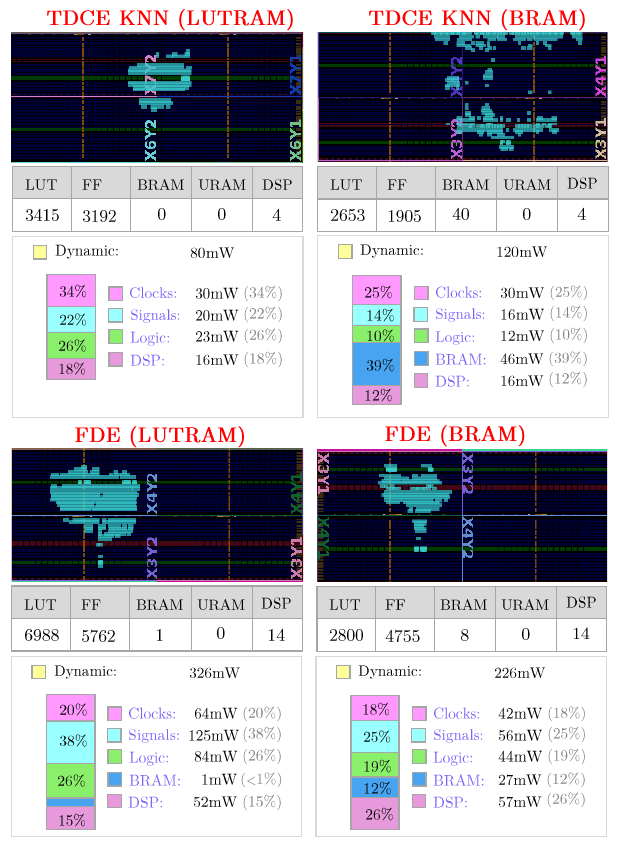}
    \put(-244,328){\textcolor{red}{\textbf{(a)}}}
    \put(-122,328){\textcolor{red}{\textbf{(b)}}}
    \put(-244,162){\textcolor{red}{\textbf{(c)}}}
    \put(-122,162){\textcolor{red}{\textbf{(d)}}}
    \caption{Memory impact comparison. FPGA implementations for 4spans (320km). a)TDCE using LUTRAM, b)TDCE using BRAM, c)FDE using LUTRAM, d)FDE using BRAM. Highlighting that for TDCE, the best memory implementation option is LUTRAM, while for FDE, the best choice is BRAM when considering energy efficiency, chip area, and routeability}
    \label{fig:memory-impact}
\end{figure}

%\J{Bigger and bigger}
Also, because in this FPGA model, the BRAMs are physically disposed of in columns~\cite{xilinx2024versal}, the chip area usage is bigger, and a bigger effort must be made by the implementation tools to route the design achieving the right timing constraints. Knowing that for this fiber length, the amount of clusters is 10 (Table~\ref{table:parameters-details}) and looking at Fig.~\ref{fig:fpga-diagram}, we conclude that each memory bank has only 10 positions, each position storing a complex number with real and imaginary parts containing 16 bits each, what gives a total of 320 bits. However, each BRAM is able to store 36~kbits \cite{xilinx2024versal}, which draws us to the conclusion that for this kind of memory, the usage efficiency is very low.  This can be explained by the fact that each BRAM has only two ports~\cite{xilinx2024versal}, and to permit the Memory Transfer block (Fig.~\ref{fig:fpga-diagram}) to copy the data from the bank memories with high speed, a larger number of memories must be used to allow simultaneous access.

In this FPGA model, BRAMs are physically disposed of in columns~\cite{xilinx2024versal}, increasing chip area usage, as shown in Fig.~\ref{fig:memory-impact}(a,b). This spreading of the chip area requires greater effort by implementation tools to meet timing constraints during routing, potentially causing timing issues in higher throughput cases that require larger areas. For the given fiber length, TDCE uses 10 clusters (Table~\ref{table:parameters-details}), resulting in each memory bank having only 10 positions, as shown in Fig.~\ref{fig:fpga-diagram}. Each position stores one complex number with 16-bit real and imaginary parts, totaling 320 bits per bank. However, each BRAM can store 36 kbits~\cite{xilinx2024versal}, resulting in very low usage efficiency. This inefficiency is due to each BRAM having only two ports~\cite{xilinx2024versal}, necessitating a larger number of memories to allow high-speed data transfer by the Memory Transfer block (Fig.~\ref{fig:fpga-diagram}) and simultaneous access.

Analyzing Fig.~\ref{fig:memory-impact}(c,d), we see that for the FDE, unlike the TDCE, dynamic power, and area usage are higher when using LUTRAM memory while the number of DSPs is not affected. As expected, the number of LUTs and FFs is greater when LUTRAM is used.

From the above analysis, we conclude that for FPGA implementation, the optimal choice in terms of chip area usage, routing, and power efficiency is to use LUTRAMs for \(x\textsubscript{S}\) memory banks in TDCE. In contrast, for the FDE, BRAMs are the better choice for the FFT/IFFT blocks. These memory implementations will be used in the following discussions.

\subsection{Results for Different Distances}

From Fig.~\ref{fig:resources}(a), we observe that the chip areas required by the GD and KNN approaches are quite similar due to their use of similar amounts of resources. Additionally, for low dispersion scenarios like 1 fiber span, the area required for the TDCE designs is smaller than that required for the FDE. However, for higher dispersion scenarios, like 8 spans, the area required is similar for all designs.

Fig.~\ref{fig:resources}(b) shows that, due to the memory implementation choices, both TDCE filters do not use BRAMs, in contrast to the FDE, which uses 8. However, the LUT quantities used by TDCE approaches are higher than those for the FDE, as explained in subsection~\ref{subsec:memory}. In addition, despite having the highest complexity in this work according to Fig.~\ref{fig:methodology-clustering-methods}(f), TDCE KNN used about one-third of the multipliers required by FDE and the same amount of multipliers as TDCE GD. This is because the multiplications for TDCE are done after the summation with overlapping between tasks (Fig.~\ref{fig:fpga-diagram}). 

To illustrate how our parallelization can save multipliers, consider the TDCE KNN for 4 spans with $M=97$, $L=20$, and $N_C=10$ (Table~\ref{table:parameters-details}). The summation process takes about $M$ clock cycles, independent of $L$ (subsection~\ref{subsec:parallelization}), plus around $N_C$ clock cycles to transfer data to secondary memory banks, totaling approximately $M+N\textsubscript{C}=107$ clock cycles, which can be rounded to 110 due to extra dataflow cycles. Each output sample requires $N\textsubscript{C}=10$ multiplications, totaling 200 multiplications that can be completed in the 110 clock cycles due to task overlapping. Thus, the multiplications can be performed with a smaller degree of parallelism (\(L\textsubscript{P}<L\)), allowing for two complex multiplications per clock cycle ($L_P=2$). Since each complex multiplication has two real multiplications, this implementation uses only 4 multipliers compared to the 14 used by FDE.

Despite having lower complexity, the GD implementation required the same number of multipliers as the KNN due to needing $N\textsubscript{C}=8$ clusters. This requires fewer parallel samples, $L=18$, totaling $L \cdot N\textsubscript{C}=144$ complex multiplications. These cannot be performed sequentially in approximately $M+N\textsubscript{C}=105$ clock cycles (TDCE GD), necessitating the same degree of parallelism as the KNN, with 2 complex multiplications per clock.

% Regarding energy efficiency, the power distribution shown in Fig.~\ref{fig:resources}(c) indicates that TDCE approaches have similar power consumption. This similarity is due to the similar quantities of required hardware and the same algorithm implementation. Both TDCE approaches exhibit lower power consumption compared to the FDE, with significant energy savings in terms of DSP and BRAM energy consumption. This effect can be attributed to the fewer multipliers required by TDCE designs and the choice to implement memories using LUTs and FFs instead of BRAMs. Additionally, energy savings on Signals are notable, explained by the simpler TDCE algorithm that avoids the need for interconnected structures like the FFT algorithm, which uses deeply interconnected "butterflies" for mathematical operations~\cite{cooley1965algorithm}. These power-distributed savings explain the superior overall energy efficiency of TDCE approaches compared to FDE, as shown in Fig.~\ref{fig:resources}(d), with an efficiency gain of up to 70.4\% even for the higher complexity TDCE KNN. It demonstrates that hardware implementation is as crucial as mathematical complexity in determining energy efficiency. It is also important to note that both TDCE designs exhibit very similar energy efficiency, despite the GD approach being less complex. This is because the reduced number of required multiplications was insufficient to translate into fewer multipliers due to the dataflow structure.

Regarding energy efficiency, Fig.~\ref{fig:resources}(c) shows that TDCE approaches have similar power consumption. This similarity is due to the similar quantities of required hardware resources and the same algorithm implementation. Both TDCE approaches consume less power than FDE, with significant energy savings in DSP and BRAM usage. This can be attributed to the fewer multipliers required by TDCE designs and the use of LUTs and FFs instead of BRAMs. Additionally, energy savings in Signals are notable, thanks to the simpler TDCE algorithm, which avoids the interconnected structures of the FFT algorithm that uses deeply interconnected "butterflies" for operations~\cite{cooley1965algorithm}. These power-distributed savings result in superior overall energy efficiency for TDCE approaches compared to FDE, as shown in Fig.~\ref{fig:resources}(d), with an efficiency gain of up to 70.4\%, even for the higher complexity TDCE KNN. This demonstrates that hardware implementation is as crucial as mathematical complexity in determining energy efficiency. Both TDCE designs exhibit very similar energy efficiency despite the GD approach being less complex. This is because the reduced number of required multiplications did not significantly reduce the number of multipliers due to the dataflow structure.

\section{Conclusion} 
\label{sec:conclusion}

% \J{which observation? You did not talk about any observation in the first sentence part}
% \J{Which mathematical properties? Too general-- specialise}

We presented a geometrical interpretation of the CDC transfer function and explained the effect of filter taps overlapping. This overlapping generates clusters of filter taps, which we harnessed to decrease the theoretical complexity of time-domain FIR filtering. Our approach was further optimized using machine learning to reduce computational complexity. We implemented the proposed solution and state-of-the-art FDE on FPGA for different fiber lengths (from 80~km to 640~km SSMF) and examined energy efficiency using the nJ/bit metric. We quantified the impact of memory implementation on resource quantity, energy efficiency, chip area usage, and routability in FPGA realization. The developed TDCE technique demonstrated up to 71.4\% savings in the number of multipliers, even for higher computational complexity cases in comparison to FDE. Furthermore, we showed that TDCE can be flexibly parallelized by leveraging the property that in linear convolution, the same filter tap multiplies consecutive input samples across consecutive output sample calculations.

In conclusion, for the considered fiber lengths, the proposed TDCE achieved up to 70.7\% energy savings compared to the FDE, better chip area usage, a slightly higher usage of LUTs, and a lower usage of FFs, despite implementing memory without BRAMs.

\section{Acknowledgments} 
\label{sec:acknowledgments}
The authors utilized ChatGPT 4 for Python, C++, and LaTeX coding support, as well as Grammarly and ChatGPT 4 for grammar correction and text clarification.

PF acknowledges \AMD{} for providing the VCK-190 board, Vivado License, and consultancy during the design of the algorithms. SKT acknowledges the EPSRC project TRANSNET (EP/R035342/1).

% % \begin{appendix}
% \begin{appendices}
% % \appendices

% \end{appendices}
% % \end{appendix}

\Urlmuskip=0mu plus 1mu\relax
\bibliographystyle{IEEEtran}
\bibliography{references}

\end{document}